\documentclass[smallextended,11pt,psfig,a4]{amsart}

\usepackage{mathrsfs}
\usepackage{amsmath}
\usepackage{amssymb}
\usepackage{braket}
\usepackage{geometry}
\usepackage{graphics}
\usepackage[pdftex]{hyperref}
\usepackage[dvips]{graphicx}
\usepackage{pgf}
\usepackage{tikz}
\usetikzlibrary{arrows,automata,positioning}

\tikzset{main node/.style={circle,draw,minimum size=1cm,inner sep=0pt},
            }

\newtheorem{theorem}{Theorem}[section]
\newtheorem{lemma}[theorem]{Lemma}
\newtheorem{pro}[theorem]{Proposition}

\newtheorem{remark}[theorem]{Remark}

\newtheorem{example}[theorem]{Example}

\begin{document}
\pagestyle{plain}



\author[ag]{Manuel D. de la Iglesia}
\address{Manuel D. de la Iglesia, Instituto de Matem\'aticas,
Universidad Nacional Aut\'onoma de M\'exico. Circuito Exterior, C.U., 04510 Ciudad de M\'exico, M\'exico.}
\email{mdi29@im.unam.mx}

\author[cfelipe]{Carlos F. Lardizabal}
\address{Carlos F. Lardizabal, Instituto de Matem\'atica e Estat\'istica, Universidade Federal do Rio Grande do Sul. Porto Alegre, RS  91509-900 Brazil.}
\email{cfelipe@mat.ufrgs.br (Corresponding author)}

\author[lv]{Newton Loebens}
\address{Newton Loebens, Instituto de Matem\'atica e Estat\'istica, Universidade Federal do Rio Grande do Sul. Porto Alegre, RS  91509-900 Brazil.}
\email{newtonloebens@gmail.com}

\newcommand\blfootnote[1]{%
  \begingroup
  \renewcommand\thefootnote{}\footnote{#1}%
  \addtocounter{footnote}{-1}%
  \endgroup
}

\def\ora{\overrightarrow}

\def\bma{\begin{bmatrix}}
\def\ema{\end{bmatrix}}
\def\bex{\begin{example}}
\def\eex{\end{example}}
\def\beq{\begin{equation}}
\def\eeq{\end{equation}}
\def\eps{\epsilon}
\def\laa{\langle}
\def\raa{\rangle}
\def\qed{\begin{flushright} $\square$ \end{flushright}}
\def\qee{\begin{flushright} $\Diamond$ \end{flushright}}
\def\ov{\overline}


\title{Quantum Markov chains on the line: matrix orthogonal polynomials, spectral measures and their statistics}

\maketitle

\begin{abstract}
Inspired by the classical spectral analysis of birth-death chains using orthogonal polynomials, we study an analogous set of constructions in the context of open quantum dynamics and related walks. In such setting, block tridiagonal matrices and matrix-valued orthogonal polynomials are the natural objects to be considered. We recall the problems of the existence of a matrix of measures or weight matrix together with concrete calculations of basic statistics of the walk, such as site recurrence and first passage time probabilities, with these notions being defined in terms of a quantum trajectories formalism. The discussion concentrates on the models of quantum Markov chains, due to S. Gudder, and on the particular class of open quantum walks, due to S. Attal et al. The folding trick for birth-death chains on the integers is revisited in this setting together with applications of the matrix-valued Stieltjes transform associated with the measures, thus extending recent results on the subject. We also consider the case of non-symmetric weight matrices and explore some examples.
\end{abstract}



\section{Introduction}

In the classical theory of Markov chains, discrete-time birth-death chains on $\mathbb{Z}_{\geq0}$ are described by a transition probability matrix of the form
\begin{equation*}\label{Pbirthinf}
P=\begin{bmatrix}
       r_0 &p_0 & 0 &0& \cdots \\
      q_1 &  r_1 & p_1 & 0&\cdots \\
      0 & q_2 &  r_2 & p_2&\cdots \\
      \vdots & \vdots & \vdots & \ddots&\ddots
    \end{bmatrix},\;\;\;r_0+p_0\leq 1,\;\;\;p_n+r_n+q_n= 1,\quad n\geq1.
\end{equation*}
Let $\{Q_n(x)\}_{n\geq 0}$ be the sequence of polynomials defined by the three-term recurrence relation
\begin{equation*}\label{3thermcod}
\begin{split}
Q_0(x)&=1,\quad Q_{-1}(x)=0, \\
xQ_n(x)&=p_nQ_{n+1}(x)+r_nQ_n(x)+q_nQ_{n-1}(x),\;\;n\geq0,
\end{split}
\end{equation*}
that is, $xQ(x)=PQ(x),$ where $Q(x)=(Q_0(x),Q_1(x),\ldots)^T.$ Then we have $x^nQ=P^nQ$, i.e.
\begin{equation}\label{compwisedisc}
x^nQ_i(x)=\sum_{k=0}^{\infty}P_{ik}^nQ_k(x),\;\;\;i\geq0.
\end{equation}
For a birth-death chain with transition probabilities $p_n,r_n,q_{n+1},n\geq0$, Favard's Theorem\index{Favard's Theorem} \cite{chih,KMc6} assures the existence of a probability measure $\psi$ supported on $[-1,1]$ such that the polynomials $\{Q_n(x)\}_{n\geq 0}$ are orthogonal with respect to $\psi$. Multiplying both sides of the equation \eqref{compwisedisc} by $Q_j(x)$  and integrating with respect to $\psi,$ we obtain the Karlin-McGregor formula \cite{KMc6}, which gives the probability of reaching vertex $j$ in $n$ steps, given that the process started at vertex $i$. This formula is given by
\begin{equation*}\label{Kmcontd}
P_{ij}^n=\frac{\displaystyle\int_{-1}^{1}x^nQ_i(x)Q_j(x)d\psi(x)}{\displaystyle\int_{-1}^{1}Q_j^2(x)d\psi(x)}.
\end{equation*}

From a theoretical point of view, it is interesting to ask whether such classical constructions can be adapted so that one can also study quantum systems as well. This has been done in the case of unitary quantum walks, where the relevant orthogonal polynomials are described in terms of the theory of CMV matrices \cite{cmv,grcpam}. Regarding the setting of open quantum dynamics \cite{benatti,davies,petruc}, the problem of obtaining orthogonal polynomials and associated measures is an interesting one as well, although we would have to consider operators which are no longer unitary.

\medskip

The main purpose of this paper is to explore the basic theory of matrix-valued orthogonal polynomials applied to an open quantum setting by providing results on weight matrices and describing several examples, hopefully encouraging the communities of quantum dynamics and orthogonal polynomials to attempt further developments on this line of research. A first step in this direction has been discussed in \cite{jl}, where a procedure for obtaining weight matrices associated with open quantum walks (OQWs) \cite{attal} on the half-line was described, this being in terms of a well-known result due to Dur\'an \cite{duran}.

\medskip

The setting we will consider in this paper concerns the class of quantum Markov chains (QMCs) on the line, as defined by S. Gudder \cite{gudder}. This model is revised in detail in Section \ref{sec2}. The main difference with OQWs is that the transition maps are not only given by conjugations of the form $X\mapsto VXV^*$, but, instead, the effect transitions can be chosen to be any  completely positive map. This larger class of examples expands the potential applicability of the theory and also makes it easier to find evolutions which are distinct from classical dynamics. 

\medskip

With an improved understanding of weight matrices, one is now able to present basic results on recurrence and positive recurrence of QMCs, as we will see in Sections \ref{sec3} and \ref{sec4}. The use of the Stieltjes transform allows us to further extend recent results on homogeneous OQWs on the line regarding criteria for site-recurrence \cite{jl20}. Sections \ref{sec5} and \ref{sec6} illustrate the theory with examples on finite segments and on the half-line, while Section \ref{sec7} explains how to consider QMCs acting on the integer line, further extending the applicability of the theory. Finally, by a proper variation of the Karlin-McGregor formula for weight matrices, we are able to discuss weight matrices which are not necessarily symmetric. This has been examined by Zygmunt \cite{zyge1,zyge2}, and such theory leads to interesting examples of QMCs, as we will see in Section \ref{sec8}.

\section{Preliminaries}\label{sec2}

Let $\mathcal{H}$ be a separable Hilbert space with inner product $\langle\,\cdot\,|\,\cdot\,\rangle$, whose closed subspaces will be referred to  as subspaces for short. The superscript ${}^*$ will denote the adjoint operator. The Banach algebra $\mathcal{B}(\mathcal{H})$ of bounded linear operators on $\mathcal{H}$ is the topological dual of its ideal $\mathcal{I}(\mathcal{H})$ of trace-class operators with trace norm
$$
\|\rho\|_1=\operatorname{\mathrm{Tr}}(|\rho|),
\qquad
|\rho|=\sqrt{\rho^*\rho},
$$
through the duality \cite[Lec. 6]{attal_lec}
\beq \label{eq:dual}
\langle \rho,X \rangle = \operatorname{\mathrm{Tr}}(\rho X),
\qquad
\rho\in\mathcal{I}(\mathcal{H}),
\qquad
X\in\mathcal{B}(\mathcal{H}).
\eeq
If $\dim\mathcal{H}=k<\infty$, then $\mathcal{B}(\mathcal{H})=\mathcal{I}(\mathcal{H})$ is identified with the set of square matrices of order $k$, denoted by $M_k(\mathbb{C})$. The duality \eqref{eq:dual} yields a useful characterization of the positivity of an operator $\rho\in\mathcal{I}(\mathcal{H})$:
\begin{equation*}\label{eq:pos-dual}
\rho\in\mathcal{I}(\mathcal{H}):
\quad
\rho\ge0 \; \Leftrightarrow \; \operatorname{\mathrm{Tr}}(\rho X)\ge0,
\quad
\forall X\in\mathcal{B}(\mathcal{H}),
\quad
X\ge0,
\end{equation*}
and similarly for the positivity of $X\in\mathcal{B}(\mathcal{H})$.

\medskip

In this paper, we assume that we have a quantum particle acting either on the integer line, the integer half-line, or on a finite segment, that is, we have that the set of vertices $V$ is labeled by $\mathbb{Z}$, $\mathbb{Z}_{\geq0}$ or a finite set $\{0,1,\dots,N\}$, respectively. In this work, vertices are also called sites. The state of the system is described by a column vector
\beq\label{qmcdens}
\rho = \begin{bmatrix} \rho_0 \\ \rho_1 \\ \rho_2 \\ \vdots \end{bmatrix},
\qquad \rho_i\in\mathcal{I}(\mathcal{H}),
\qquad \rho_i\ge0,
\qquad \sum_{i\in V}\operatorname{\mathrm{Tr}}(\rho_i)=1.
\eeq
After one time step, the system evolves to the state $\Phi(\rho)$ given by $\Phi(\rho)_i=\sum_{j\in V}\Phi_{ij}(\rho_j)$, where
$$
\Phi =
\begin{bmatrix}
\\[-10pt]
\Phi_{00} & \Phi_{01} & \Phi_{02} & \dots
\\[2pt]
\Phi_{10} & \Phi_{11} & \Phi_{12} & \dots
\\[2pt]
\Phi_{20} & \Phi_{21} & \Phi_{22} & \dots
\\
\dots & \dots & \dots & \dots
\end{bmatrix},
$$
is called a {\bf Quantum Markov Chain} (QMC) \cite{gudder}: this means that the $\Phi_{ij}$ are completely positive (CP) maps on $\mathcal{I}(\mathcal{H})$ and the column sums $\sum_{i\in V} \Phi_{ij}$ are trace-preserving (TP) (the summations are assumed to converge in the strong operator topology), see Figure 1. A density $\rho$ of the form \eqref{qmcdens} will be called a {\bf QMC density}. {\color{black}The set of density operators acting on a subspace $\mathcal{K}$ of $\mathcal{H}$ will be denoted by $\mathcal{D}(\mathcal{K}).$
}

\medskip

An important particular class of CP maps is given by the ones of the form
\beq\label{def_oqw}
\Phi_{ij}(\rho)=B_{ij}\rho B_{ij}^*,
\qquad
B_{ij}\in \mathcal{B}(\mathcal{H}),
\qquad
\sum_{k \in V} B_{kj}^*B_{kj}=I,
\qquad
\forall\; i,j \in V.
\eeq
The summation above must be understood in the strong sense, and the corresponding identity is the trace- preserving condition for the columns of the QMC $\Phi$. We will say that $B_{ij}$ is the effect matrix of transitioning from vertex $j$ to vertex $i$. QMCs for which $\Phi_{ij}$ can be written in the form \eqref{def_oqw} are called {\bf Open Quantum Random Walks} (OQWs), following the terminology established by S. Attal et al. \cite{attal}. Explicitly, OQWs are QMCs of the form
\beq\label{eq:OQW}
\Phi(\rho) =
\sum_{i\in V}\left(\sum_{j\in V}
B_{ij}\rho_j B_{ij}^*\right)\otimes |i\rangle\langle i|,
\eeq
and, as any QMC, they may be alternatively seen as CP-TP maps on $\mathcal{I}(\mathcal{H}\otimes V)$.

\medskip

\begin{center}
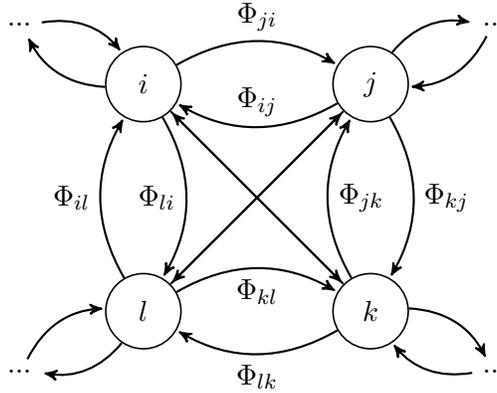
\begin{figure}[ht]
\begin{tikzpicture}
[->,>=stealth',shorten >=1pt,auto,node distance=2.0cm,
                    semithick]
    \node[main node] (1) {$i$};
    \node[main node] (2) [right=2.0cm and 2.0cm of 1,label={[]$$}]  {$j$};

    \node[main node] (3) [below=2.0cm and 2.0cm of 2,label={[]$$}]  {$k$};
    \node[main node] (4) [left = 2.0cm and 2.0cm of 3] {$l$};

    \node[] (1a) [left=1.0cm and  1.0cm of 1] {};
    \node[] (1b) [above=0.5cm and  0.5cm of 1a] {$...$};

    \node[] (2a) [right=1.0cm and  1.0cm of 2] {};
    \node[] (2b) [above=0.5cm and  0.5cm of 2a] {$...$};

    \node[] (3a) [right=1.0cm and  1.0cm of 3] {};
    \node[] (3b) [below=0.5cm and  0.5cm of 3a] {$...$};

    \node[] (4a) [left=1.0cm and  1.0cm of 4] {};
    \node[] (4b) [below=0.5cm and  0.5cm of 4a] {$...$};



    \path[draw,thick]


    (1) edge   [bend left]  node {$\Phi_{ji}$}    (2)
    (1) edge   [bend left]  node[above]{$$}   (1b)
    (1b) edge   [bend left]  node[below] {$$}    (1)

    (2) edge   [bend left]     node[above]{$\Phi_{ij}$} (1)
    (2) edge   [bend left]  node[above]{$$}   (2b)
    (2b) edge   [bend left]  node[below] {$$}    (2)

    (3) edge   [bend left]  node[above]{$$}   (3b)
    (3b) edge   [bend left]  node[below] {$$}    (3)

    (4) edge   [bend left]  node[above]{$$}   (4b)
    (4b) edge   [bend left]  node[below] {$$}    (4)

    (2) edge   [bend left]     node {$\Phi_{kj}$} (3)
    (3) edge   [bend left]     node[right] {$\Phi_{jk}$} (2)

    (1) edge   [bend left]     node[left] {$\Phi_{li}$} (4)
    (4) edge   [bend left]     node[left] {$\Phi_{il}$} (1)

    (4) edge   [bend left]     node[below] {$\Phi_{kl}$} (3)
    (3) edge   [bend left]     node[below] {$\Phi_{lk}$} (4)


    (3) edge      node {$$} (1)
    (1) edge        node {$$} (3)

    (2) edge      node {$$} (4)
    (4) edge        node {$$} (2)



    ;
\end{tikzpicture}
\caption{Schematic illustration of QMCs. The walk is realized on a graph with a set of vertices denoted by $i,j,k,l,\dots$ and each operator $\Phi_{ij}$ is a completely positive map describing a transformation in the internal degree of freedom of the particle during the transition from vertex $j$ to vertex $i$. For simplicity of illustration some edges are not labeled. In the particular case that all maps are conjugations, i.e., for every $i,j$, $\Phi_{ij}=B_{ij}\cdot B_{ij}^*$ for certain matrices $B_{ij}$ the QMC is called an open quantum walk. In this work, the graphs considered will be either a line segment, the half-line, or the integer line.}\label{fig0}
\end{figure}
\end{center}

The vector representation $vec(A)$ of $A\in M_k(\mathbb{C})$, given by stacking together its rows, will be a useful tool. For instance,
\begin{equation*}
A = \begin{bmatrix} a_{11} & a_{12} \\ a_{21} & a_{22} \end{bmatrix}
\quad\Rightarrow\quad
vec(A):=\begin{bmatrix} a_{11} \\ a_{12} \\ a_{21} \\ a_{22}\end{bmatrix}.
\end{equation*}
The $vec$ mapping satisfies $vec(AXB^T)=(A\otimes B)\,vec(X)$ {\cite{hj2}} for any square matrices $A, B, X$, with $\otimes$ denoting the Kronecker product. In particular, $vec(BXB^*)=vec(BX\ov{B}^T)=(B\otimes \ov{B})\,vec(X)$,
from which we can obtain the matrix representation $\widehat\Phi$ for a CP map $\sum_i B_i\cdot B_i^*$ when the underlying Hilbert space $\mathcal{H}$ is finite-dimensional:
\begin{equation*}\label{matrep}
\widehat\Phi = \sum_{i} \lceil B_{i} \rceil,
\qquad \lceil B \rceil := B \otimes \ov{B}.
\end{equation*}
Here the operators $B_i$ are identified with some matrix representation. We have that $\lceil B \rceil^* = \lceil B^*\rceil $, where $B^*$ denotes the Hermitian transpose of a matrix $B$. Then, the vector and matrix representation of states and CP maps may be easily adapted to QMCs. In fact, since any element of $\mathcal{I}_V(\mathcal{H})$ is block diagonal, when $\dim\mathcal{H}<\infty$, it may be represented by combining the vector representations of the finite diagonal blocks,
$$
\rho=\sum_{i\in V} \rho_i\otimes|i\rangle\langle i|
\quad\Rightarrow\quad
\overrightarrow{\rho}:=\begin{bmatrix} vec(\rho_1) \\ vec(\rho_2) \\ \vdots \end{bmatrix}.
$$
Then, the OQW \eqref{eq:OQW} admits the block matrix representation
\begin{equation*}\label{eq:mrOQW}
\overrightarrow{\Phi(\rho)} =
\widehat{\Phi}\,\overrightarrow{\rho},
\qquad
\widehat{\Phi} =
\begin{bmatrix}
\lceil B_{00} \rceil & \lceil B_{01} \rceil & \cdots
\\[2pt]
\lceil B_{10} \rceil & \lceil B_{11} \rceil & \cdots
\\
\vdots & \vdots
\end{bmatrix},
\end{equation*}
and analogously for QMCs. We will often identify $\Phi$ with its block matrix representation and omit the hat, as the usage of such object will be clear from the context. Also, we will sometimes write $X$ instead of $\lceil X\rceil$ in contexts where no confusion arises.

\medskip

Although the above definitions concern QMCs on general graphs, we remark that in this paper we will deal exclusively with the one-dimensional situation, more specifically, with the  nearest neighbor QMC or quantum birth-death chain, e.g.,
\begin{equation}\label{tpm}
{\Phi}=
\begin{bmatrix}
  B_0 & C_1 & && \\
  A_0& B_1 & C_2  && \\
 &A_1& B_2 & C_3  & \\
  &  & \ddots  &\ddots&\ddots
\end{bmatrix},
\end{equation}
for certain operators $A_i, B_i, C_i$, and the remaining ones being equal to zero.

\subsection{The calculation of probabilities for QMCs} By letting $\rho\otimes|i\rangle\langle i|$  be an initial density matrix concentrated at site $|i\rangle$, we can describe $n$ iterations of the QMC (\ref{tpm}). By setting $\rho^{(0)}=\rho\otimes|i\rangle\langle i|$, $\mathrm{Tr}(\rho)=1$, we write (assume $C_0=0$)
$$
\Phi^{n}(\rho\otimes|i\rangle\langle i|)=\sum_{k\geq 0} \rho_k^{(n)}\otimes|k\rangle\langle k|,\;\;\;\rho_k^{(n)}=C_k\rho_{k+1}^{(n-1)}C_k^*+B_k\rho_{k}^{(n-1)}B_k^*+A_k\rho_{k-1}^{(n-1)}A_k^*, \;\;\;n=1,2,\dots
$$
Then, the probability of reaching site $|j\rangle$ at the $n$-th step, given that we started at site $|i\rangle$ with initial density $\rho$ concentrated at $i$ is given by
\begin{equation*}\label{oqwprobmat}
p_{ji;\rho}(n)=p_{n}(\rho\otimes |i\rangle \to |j\rangle):=\mathrm{\mathrm{Tr}}(\rho_j^{(n)})=\mathrm{\mathrm{Tr}}\left(\mathrm{vec}^{-1}\left[(\widehat{\Phi}^n)_{ji}\mathrm{vec}(\rho)\right]\right),
\end{equation*}
where $(\widehat{\Phi}^n)_{ji}$ is the $(j,i)$-th block of the block matrix $\widehat{\Phi}^n$, the $n$-th power of the block representation $\widehat{\Phi}$.

Following \cite{bardet,cgl}, we say that vertex $i$ is {\bf recurrent} with respect to $\rho$, or simply {\bf $\rho$-recurrent}, if
$$\sum_{n=0}^\infty  p_{ii;\rho}(n)=\infty.$$
Otherwise, we say that vertex $i$ is {\bf transient} with respect to $\rho$, or {\bf $\rho$-transient}. We say that, with respect to a fixed QMC, vertex $i$ is {\bf recurrent} if it is $\rho$-recurrent with respect to every density $\rho$ concentrated in $i$, and {\bf transient} if it is $\rho$-transient with respect to every density in $i$. Finally, we say that a QMC $\Phi$ is recurrent if every site is recurrent, and we define transient QMCs analogously.

\begin{remark} We note that in the setting of QMCs, one can also consider the notion of monitored recurrence, see e.g.  \cite{bardet,glv,jl20}. For simplicity, we will not consider such definition in this work, and we refer the reader to the references for a detailed discussion on such matter.\end{remark}

Finally, we will be able to discuss expected return times to sites of QMCs in terms of the following notion. Let $T$ denote a positive map (that is, such that if $X\geq 0$ then $T(X)\geq 0$) acting on the space $\mathcal{I}(\mathcal{H})$ of trace-class operators of a Hilbert space $\mathcal{H}$. We say that $T$ is irreducible if the only orthogonal projections $P$ such that $T(P \mathcal{I}(\mathcal{H})P)\subset P \mathcal{I}(\mathcal{H})P$, are $P=0$ and $P=I$, see \cite{carbone1,carbone2} for more on this. Then,  we say that a QMC $\Phi$ is {\bf positive recurrent} if it is irreducible and if it admits an invariant distribution. We note that by [\cite{bardet}, Thm. 4.3 and 4.5] for positive recurrent OQWs, we have finite expected return times for every density and site, and the same reasoning provides the analogous result in the case of QMCs.

\subsection{Auxilliary notation: compact form} In some of the examples we study in this paper we will use the following algebraic simplification. We know that the matrix representation of the conjugation map induced by an order 2 matrix $M=(m_{ij})$ is given by
$$\lceil M\rceil=M\otimes \ov{M}=\begin{bmatrix} |m_{11}|^2 & m_{11}\ov{m_{12}} & \ov{m_{11}}m_{12} & |m_{12}|^2 \\ m_{11}\ov{m_{21}} & m_{11}\ov{m_{22}} & m_{12}\ov{m_{21}} & m_{12}\ov{m_{22}} \\
\ov{m_{11}}m_{21} & \ov{m_{12}}m_{21} & \ov{m_{11}}m_{22} & \ov{m_{12}}m_{22} \\ |m_{21}|^2 & m_{21}\ov{m_{22}} & \ov{m_{21}}m_{22} & |m_{22}|^2\end{bmatrix}=\begin{bmatrix} a& b & \ov{b} & c \\ d & e & f & g \\ \ov{d} & \ov{f} & \ov{e} & \ov{g} \\ h & j & \ov{j} & k\end{bmatrix},\;\;\;m_{ij}\in\mathbb{C}.$$
Let us consider the setting for which all of the above coefficients are real, and acting on positive semidefinite matrices with real entries. Then
$$
\lceil M\rceil vec(\rho)=\begin{bmatrix} a& b & b & c \\ d & e & f & g \\ d & f & e & g \\ h & j & j & k\end{bmatrix}\begin{bmatrix} x \\ y \\ y \\ z\end{bmatrix}=\begin{bmatrix} ax+2by+cz \\ dx+(e+f)y+gz \\dx+(e+f)y+gz \\ hx+2jy+kz\end{bmatrix},\;\;\;\rho=\begin{bmatrix} x & y \\ y & z\end{bmatrix}.
$$
In this particular setting we note that the above computation can be codified in a more economic way, namely,
via the correspondence
\begin{equation}\label{compform}
\lceil M\rceil vec(\rho)  \;\leftrightarrow \; \check{M}\check{\rho}:=\begin{bmatrix} a & 2b & c \\ d & e+f & g \\ h & 2j & k\end{bmatrix}\begin{bmatrix} x \\ y \\ z\end{bmatrix}=\begin{bmatrix} a+2by+cz \\ dx+(e+f)y+gz \\ hx+2jy+kz \end{bmatrix}.
\end{equation}
We call the map $\check{M}$ the {\bf compact form} of the conjugation induced by $M$, or simply the compact form of $M$. It is clear that many calculations coming from quantum mechanical models can be written in terms of real numbers only and, even though the real coefficient assumption often precludes us from complete generality, we are still able to learn useful information about 1-qubit quantum channels.

\medskip

The following properties of the compact form are proven by a routine calculation:
\begin{enumerate}
\item  $\check{}(MR)=\check{M}\check{R}$ for any matrices, resembling the matrix representation property $\lceil MR\rceil=\lceil M\rceil\lceil R\rceil$.
\item The compact form preserves the computation of product of conjugations acting on positive definite matrices.  That is, if $M$ and $R$ are matrices then $\lceil M\rceil\lceil R\rceil vec(\rho)$ corresponds to $\check{M}\check{R}\check{\rho}$.
\end{enumerate}

\section{Weight matrices}\label{sec3}

Let $W$ be a weight matrix, i.e. a $N\times N$ matrix of measures supported in the real line such that $dW(y)- dW(x)\geq 0$ (positive semidefinite) for $x<y$. We also allow the case of discrete measures, those appearing naturally in the case of walks acting on a finite number of vertices. Define the matrix-valued inner product given by
\begin{equation}\label{innprod}
(P,Q):=\int_\mathbb{R} P^*(x)dW(x)Q(x).
\end{equation}
Also regarding positive semidefiniteness, we recall that $(P,P)\geq 0$, $(P,P)>0$ whenever det$(P)\not\equiv 0$ and $(P,P)=0$ if and only if $P\equiv 0$. Let $\{Q_n(x)\}_{n\geq0}$ denote a sequence of matrix-valued orthogonal polynomials with respect to such product, with nonsingular leading coefficients. Then
$$
\int_{\mathbb{R}} Q_n^*(x)dW(x)Q_m(x)=\|Q_n\|^2\delta_{nm}.
$$
The set of polynomials will be called orthonormal if $\|Q_n\|^2=(Q_n,Q_n)=I, n\geq0$. It is well-known that any family of matrix-valued orthogonal polynomials satisfies a three-term recurrence relation of the form
\begin{equation}\label{polrec}
xQ_n(x)=Q_{n+1}(x)A_n+Q_n(x)B_n+Q_{n-1}(x)C_n,\;\;n\geq0,\quad Q_0(x)=I,\quad Q_{-1}(x)=0,
\end{equation}
for certain $A_n, B_n,C_{n+1}, n\geq0,$ square matrices. This gives rise to a block tridiagonal Jacobi matrix of the form
\beq\label{tridiag1}
{P}=\begin{bmatrix} B_0 & C_1 & & & 0 \\
                    A_0 & B_1 & C_2 &  & & \\
                     & A_1 & B_2 & C_3 & \\
                      0& & \ddots & \ddots & \ddots\end{bmatrix},
\eeq
so that (\ref{polrec}) can be written as $xQ(x)=Q(x)P$, where $Q(x)=(Q_0(x),Q_1(x),\dots)$. Let us now see the inverse problem, i.e. under what conditions we can guarantee the existence of a weight matrix given a block tridiagonal matrix of the form \eqref{tridiag1}. As discussed previously, namely, whenever the weight matrix exists, the $(i,j)$-th block of the block matrix ${P}^n$ can be written as
\begin{equation*}\label{kmcgmatrix}
({P}^n)_{ij}=(Q_i(x),Q_i(x))^{-1}\left(\int_{\mathbb{R}} x^nQ_i^*(x)dW(x)Q_j(x)\right).
\end{equation*}
However, unlike the one-dimensional case, a system of matrix-valued polynomials $\{Q_n(x)\}_{n\geq 0}$ satisfying such recurrence relation  is not necessarily orthogonal with respect to an inner product induced by a weight matrix. In view of this, Dette et al. describe an existence criterion:
\begin{theorem}(\cite[Theorem 2.1]{dette})\label{dette 2.1} Assume that the matrices $A_n,C_{n+1},n\geq0,$ in the one-step block  tridiagonal transition matrix \eqref{tridiag1} are nonsingular. There exists a weight matrix $W$ supported on the real line such that the polynomials defined by \eqref{polrec} are orthogonal with respect to the measure $dW(x)$ if and only if  there exists a sequence of nonsingular matrices $\{R_n\}_{n\geq0}$ such that
\label{CondiDette}
\begin{enumerate}
\item $R_nB_nR_n^{-1} \textrm{ is Hermitian, }\; \forall\; n=0,1,2,\dots$.
\item $R_n^*R_n=\left(A_0^*\cdots A_{n-1}^*\right)^{-1}(R_0^*R_0)C_1\cdots C_n,\;\;\forall\; n=1,2,\dots$.
\end{enumerate}
\end{theorem}
In the case of a QMC with block tridiagonal matrix of the form
\begin{equation}\label{oqwbas1}
\widehat{\Phi}=\begin{bmatrix} \lceil B_0\rceil & \lceil C_1\rceil & & & 0 \\
                    \lceil A_0\rceil & \lceil B_1\rceil  & \lceil C_2\rceil &  & & \\
                     & \lceil A_1\rceil & \lceil B_2\rceil  & \lceil C_3\rceil & \\
                      0& & \ddots & \ddots & \ddots\end{bmatrix},
\end{equation}
then, in order to find the corresponding weight matrix, we need to find nonsingular matrices $\{R_n\}_{n\geq0}$ such that
$$
\Pi_n:=R_n^*R_n=(\lceil A_0\rceil^*\cdots \lceil A_{n-1}\rceil^*)^{-1}\Pi_0\lceil C_1\rceil\cdots \lceil C_{n}\rceil \;\; \text{ and } \;\; \Pi_n\lceil B_n\rceil=\lceil B_n\rceil^*\Pi_n,\;\;\;n=1,2,\dots
$$

Finally, we note that we have a version of the Karlin-McGregor formula for QMCs, in close analogy with the result seen in \cite[Theorem 2.1]{jl}:
\begin{theorem}\label{kmcgt}(Karlin-McGregor formula for QMCs).  Let $\widehat{\Phi}$ in \eqref{oqwbas1} be the matrix representation of a QMC $\Phi$. Assume that there exists a weight matrix $W$ associated with $\widehat{\Phi}$. Then we have
\begin{equation*}
p_{ji;\rho}(n)=\mathrm{\mathrm{Tr}}\left(\mathrm{vec}^{-1}\left[(Q_j(x),Q_j(x))^{-1}\left(\int_{\mathbb{R}}x^nQ_j^*(x)dW(x)Q_i(x)\right)\mathrm{vec}(\rho)\right]\right),
\end{equation*}
where $\rho=\rho_i\otimes|i\rangle\langle i|$ is a density matrix concentrated on vertex $i$ and $\{Q_n(x)\}_{n\geq0}$ are the matrix-valued orthogonal polynomials defined by \eqref{polrec}.
\end{theorem}
\begin{remark}
The inner product introduced in \eqref{innprod} is different from the one used in many papers on this subject (see for instance \cite{dette,duran,grunb,jl,zyge1,zyge2} and references therein). The standard inner product used is called \textbf{left} inner product
\begin{equation*}
(P,Q)_L:=\int_\mathbb{R} P(x)dW(x)Q^*(x),
\end{equation*}
which is different from the one defined by \eqref{innprod}, which sometimes is called \textbf{right} inner product (see \cite{SvA}). We obviously have $(P,Q)=(P^*,Q^*)_L$.
\end{remark}

\section{Recurrence and first passage time probabilities}\label{sec4}

Consider the Stieltjes transform of a weight matrix $W$ with support on the real line given by
\begin{equation}\label{stieltjesDef}
B(z;W):=\int_\mathbb{R} \dfrac{dW(x)}{z-x},\;\;\;z\in \mathbb{C}\backslash\mathbb{R}.
\end{equation}
Let $N\in\{1,2,\ldots\}$ and $\Phi$ be a QMC described by
\begin{equation}\label{phirenewaleq}
{\Phi}=
\begin{bmatrix}
  B_0 & C_1 & && \\
  A_0& B_1 & C_2  && \\
 &A_1& B_2 & C_3  & \\
  &  & \ddots  &\ddots&\ddots
\end{bmatrix},
\end{equation}
where $A_n,B_n,C_{n+1}\in M_{N^2}(\mathbb{C}),\;n\geq 0$.  Assume there exists a weight matrix $W$ such that
\begin{equation}\label{Kmc44}
\Phi_{ij}^{(n)}=\Pi_i\left(\int_{\mathbb{R}}x^nQ_i^*(x)dW(x)Q_j(x)\right),
\end{equation}
where $\Pi_i=\left(\int_{\mathbb{R}}Q_i^*(x)dW(x)Q_i(x)\right)^{-1}$. Now let us define a generating function associated with hitting probabilities from $j$ to $i$ with respect to the QMC $\Phi$, i.e.
\begin{equation}\label{genfunct}
\Phi_{ij}(s):=\sum_{n=0}^\infty \Phi_{ij}^{(n)}s^n,\;\;\;\Phi_{ij}^{(n)}=\mathbb{P}_i\Phi^n\mathbb{P}_j,
\end{equation}
where $\mathbb{P}_k$ is the projection map onto the space generated by the state $\ket{k}$ on $\mathbb{Z}_{\geq0}$. We will start with the following result concerning $\rho$-recurrence.
\begin{theorem}\label{OQWteostieltjes}
Let $\rho$ be some density. A vertex $i\in V$ is $\rho$-recurrent if and only if
\begin{equation*}\label{stieltjesteorec}
\lim_{s\uparrow 1}\mathrm{Tr}\left[vec^{-1}\left(\Pi_i\int_\mathbb{R} \frac{1}{1-sx}Q_i^*(x)dW(x)Q_i(x)vec(\rho)\right)\right]=\infty.
\end{equation*}
As a consequence, vertex $\ket{0}$ is $\rho$-recurrent if and only if
\begin{equation}\label{corqmcstieltjes}
\lim_{z\downarrow 1}\mathrm{Tr}\left[\; vec^{-1}\left(B(z;W)vec(\rho)\right)\right]=\infty,
\end{equation}
where $B(z;W)$ is defined by \eqref{stieltjesDef}.
\begin{proof}
By Fubini's Theorem and for $|sx|<\infty$ we have
\begin{equation}\label{fubstieltjes}
\begin{split}
\Phi_{ji}(s) &=\sum_{n=0}^{\infty}s^n\Phi_{ji}^{(n)}= \sum_{n=0}^{\infty}\Pi_j \int_\mathbb{R} (sx)^nQ_j^*(x)dW(x)Q_i(x)\\
&=\Pi_j \int_\mathbb{R}  \sum_{n=0}^{\infty}(sx)^nQ_j^*(x)dW(x)Q_i(x)=\Pi_j\int_\mathbb{R}  \frac{1}{1-sx}Q_j^*(x)dW(x)Q_i(x).
\end{split}
\end{equation}
Then
$$
\lim_{s\uparrow 1}\mathrm{Tr}\left(vec^{-1}\left(\Phi_{ji}(s)vec(\rho)\right)\right)=\lim_{s\uparrow 1}\sum_{n=0}^{\infty}\mathrm{Tr}\left(vec^{-1}\left(s^n\Phi_{ji}^{(n)}vec(\rho)\right)\right)=\sum_{n=0}^{\infty}p_{ji;\rho}(n).
$$
By taking $s=1/z$, we obtain \eqref{corqmcstieltjes}.
\end{proof}
\end{theorem}
In a similar way we can prove that an irreducible QMC $\Phi$ with associated weight matrix $W$ is recurrent with respect to some density $\rho$ if and only if
$$
\lim_{s\uparrow 1}\mathrm{\mathrm{Tr}}\left(\int \frac{dW(x)}{1-xs}\rho\right)=\infty.
$$

Regarding positive recurrence in terms of the spectral matrix $W$, we have the following:
\begin{pro} For an irreducible QMC $\Phi$ \eqref{phirenewaleq} admitting a weight matrix $W$, the walk is positive recurrent if and only if the weight matrix $W$ has a finite jump at $x=1$.
\end{pro}
{\bf Proof.} An irreducible, positive recurrent QMC always admits a faithful (strictly positive), invariant distribution by \cite[Theorem 5.8]{ls2015}. Therefore, we conclude, by \cite[Corollary 5.4]{carbone1}, that
$$\lim_{n\to\infty}\mathrm{\mathrm{Tr}}(\mathbb{P}_0\Phi^{2n}\mathbb{P}_0\rho)>0.$$
Since $x^{2n}\to0$ monotonically in $x\in (-1, 1)$, from Theorem \ref{kmcgt} we see that the limit is positive if the spectral measure has positive jumps at $x = 1$ or at $x = -1$. However, there cannot be a jump at
$x = -1$ since, otherwise, the size of the jump would be
$$-\lim_{n\to\infty}\mathrm{\mathrm{Tr}}\left(\mathrm{vec}^{-1}\left[\int_{-1}^1 x^{2n+1}dW(x)\mathrm{vec}(\rho)\right]\right)=-\lim_{n\to\infty}\mathrm{\mathrm{Tr}}(\mathbb{P}_0\Phi^{2n+1}\mathbb{P}_0\rho)\leq 0.$$
But this quantity must be positive, so there is no jump at $x = -1$, for any choice of density $\rho$. Therefore, the QMC is positive recurrent if and only if there is a jump at $x = 1$.
\qed
Let us now derive an expression for first passage probabilities of QMCs in terms of matrix-valued polynomials only. The following discussion is inspired by the classical reasoning presented in \cite{dIbook}, with the main result being formula \eqref{f=QQ} presented below, which allows us to obtain first visit probabilities in terms of matrix polynomials in a simple manner. For $k\geq 0,$ consider the  QMC $\Phi$ with matrix representation
\begin{equation*}\label{kthdoqw}
\Phi=\left[\begin{array}{cccc|cccc}
  B_0 & C_1 &  &  &  &  &  &  \\
  A_0 & B_1 & C_2 &  &  &  &  &  \\
 & \ddots & \ddots & \ddots &  &  &  &  \\
  &  & A_{k-1} & B_k &C_{k+1} &  &  &  \\
  \hline
   &  &  & A_{k} & B_{k+1} & C_{k+2} &  &  \\
    &  &  &  & A_{k+1} & B_{k+2} & C_{k+3} &  \\
     &  &  &  &  & \ddots & \ddots &\ddots
\end{array}\right],
\end{equation*}
where $B_n,A_n,C_{n+1}\in M_N(\mathbb{C}),\;n\geq 0$. As usual, we recursively define the following matrix-valued polynomials,
\begin{equation}\label{3thermmatrix}
\begin{split}
Q_0(x)&=I_N,\quad Q_{-1}(x)=0 \\
  xQ_n(x) &= Q_{n+1}(x)A_n+Q_n(x)B_n+Q_{n-1}(x)C_n,
\end{split}
\end{equation}
that is, $xQ(x)=Q(x)\Phi,$ where $Q(x)=(Q_0(x),Q_1(x),\ldots).$ Suppose that $\Phi$ satisfies the conditions of Theorem \ref{dette 2.1}, so the polynomials defined by \eqref{3thermmatrix} are orthogonal with respect to a weight matrix $W$ and $\Pi \Phi=\Phi^*\Pi$, where $\Pi=\mbox{diag}(\Pi_0,\Pi_1,\ldots)$ and $\Pi_{j}=R_j^*R_j,j\geq0$. Analogously to the classical case, we define the {\bf $k$-th associated polynomials}
\begin{equation*}\label{kth pols doqw}
xQ_n^{(k)}(x)=\delta_{nk}+Q_{n+1}^{(k)}(x)A_n+Q_n^{(k)}(x)B_n+Q_{n-1}^{(k)}(x)C_n.
\end{equation*}
Note that $Q_n^{(k)}(x)=0$ if $0\leq n\leq k$ and deg$(Q_n^{(k)}(x))=n-k-1$ if $n>k.$ Consider the generating function $\Phi(s)$ associated with $\Phi$ defined by \eqref{genfunct}. Assuming $\|s\Phi\|<1,$ $\Phi_{ji}(s)$ converges for every $i,j,$ thus
$$
\sum_{n=0}^{\infty}(s\Phi)^n(I-s\Phi)=I\Rightarrow \Phi(s)-\Phi(s)(s\Phi)=I.
$$
Therefore, we have the equation
\begin{equation*}\label{IsPhi}
\Phi(s)=I+\Phi(s)(s\Phi),
\end{equation*}
which can be rewritten by blocks as
\begin{equation}\label{ijPhiblocks}
\begin{split}
\Phi_{j0}(s)  &= \delta_{j0}+\Phi_{j0}(s)B_0+\Phi_{j1}(s)A_0,\quad j\geq0 \\
\Phi_{ji}(s)  &= \delta_{ji}+\Phi_{j,i-1}(s)C_i+\Phi_{j,i}(s)B_i+\Phi_{j,i+1}(s)A_i,\quad i\ge1,j\ge0.
\end{split}
\end{equation}
A particular solution of \eqref{ijPhiblocks}  is given by
$$
\Phi_{ji}(s)=s^{-1}Q_i^{(j)}(s^{-1}).
$$
On the other hand, the general solution of $\Phi(s)=\Phi(s)(s\Phi),$ which is
$$
\Phi_{ji}(s)=g_j(s)Q_i(s^{-1})
$$
gives
\begin{equation*}\label{gip}
\Phi_{ji}(s)=\Phi_{j,i-1}(s)C_i+\Phi_{j,i}(s)B_i+\Phi_{j,i+1}(s)A_i,
\end{equation*}
and consequently, the general solution of \eqref{ijPhiblocks} is
$$
\Phi_{ji}(s)=s^{-1}Q_i^{(j)}(s^{-1})+g_j(s)Q_i(s^{-1}).
$$
Since $Q_0^{(j)}=0$ and $Q_0=1,$ one has $\Phi_{j0}(s)=g_j(s)Q_0(s^{-1})=g_j(s).$ Moreover, since $\Phi_{ji}^{(n)}=\Pi_j^{-1}\Phi_{ij}^{(n)*}\Pi_i,$ we have
$$\Phi_{j0}(s)=\sum_{n=0}^{\infty}s^n\Pi_j^{-1}\Phi_{0j}^{(n)*}\Pi_0=\Pi_j^{-1}\Phi_{0j}(s)^*\Pi_0,
$$
so we obtain the general solution for $g_j(s):$
\begin{align*}
g_j(s) &= \Phi_{j0}(s)=\Pi_j^{-1}\Phi_{j0}(s)^*\Pi_0\\
&=\Pi_j^{-1}\left(s^{-1}Q_j^{(0)}(s^{-1})+g_0(s)Q_j(s^{-1})\right)^*\Pi_0= \Pi_j^{-1}\left(s^{-1}Q_j^{(0)}(s^{-1})+\Phi_{00}(s)Q_j(s^{-1})\right)^*\Pi_0.
\end{align*}
Therefore the general solution for $\Phi_{ij}(s)$ is given by
\begin{equation}\label{solPhiz}
\Phi_{ji}(s)=s^{-1}Q_i^{(j)}(s^{-1})+\Pi_j^{-1}\left(s^{-1}Q_j^{(0)}(s^{-1})+\Phi_{00}(s)Q_j(s^{-1})\right)^*\Pi_0Q_i(s^{-1}).
\end{equation}
If we assume $i< j,$ then $Q_i^{(j)}=0$ and \eqref{solPhiz} becomes
\begin{equation}\label{solPhizbec}
\Phi_{ji}(s)=\Pi_j^{-1}\left(s^{-1}Q_j^{(0)}(s^{-1})+\Phi_{00}(s)Q_j(s^{-1})\right)^*\Pi_0Q_i(s^{-1}).
\end{equation}
Now consider the first passage time operator $F(s)$ satisfying
\begin{equation}\label{FvsPhidef}
\begin{split}
F(s) &= [F_{ji}(s)]_{j,i=0,1,2,\ldots} \\
  F_{ji}(s) &= \Phi_{jj}(s)^{-1}(\Phi_{ji}(s)-\delta_{ji}I),
\end{split}
\end{equation}
that is, with definition given by
\begin{equation}\label{firstVFdoqw2}
F(z)=z\mathbb{P}\Phi(I-z\mathbb{Q}\Phi)^{-1},
\end{equation}
where $\mathbb{P}$ and $\mathbb{Q}=I-\mathbb{P}$ are bounded projections from $\mathcal{H}$ onto supplementary closed subspaces of $\mathcal{H}.$ Further, we denote by $\mathbb{P}_k$ the projection map onto the space generated by the state $\ket{k}$ on $\mathbb{Z}_{\geq0}$ and $\mathbb{Q}_k:=I-\mathbb{P}_k$. In this way, we are able to calculate the probability of every reaching vertex $j$, given that we have started at vertex $i$ and density $\rho$, by writing
$$p(\rho\otimes |i\rangle \to|j\rangle)={\color{black}\lim_{z\uparrow 1}\textmd{Tr}\left(F_{ji}(z)\rho\right)}={\color{black}\lim_{z\uparrow 1}\textmd{Tr}\left(z\mathbb{P}_j\Phi(I-z\mathbb{Q}_j\Phi)^{-1}\rho\right)}.$$

\medskip

By \cite{gv}, $F(s)$ defined as above indeed satisfies equation \eqref{FvsPhidef}. So, let $i< j$ and
$\rho\in M_N(\mathbb{C}),$ then by equation \eqref{solPhizbec}
\begin{equation*}
\begin{split}
F_{ji}(s)&= \Phi_{jj}(s)^{-1}\Phi_{ji}(s)\\
&= Q_j(s^{-1})^{-1}\Pi_0^{-1}\left[\left(s^{-1}Q_j^{(0)}(s^{-1})+\Phi_{00}(s)Q_j(s^{-1})\right)^*\right]^{-1}\Pi_j \\
&\quad\times\Pi_j^{-1}\left(s^{-1}Q_j^{(0)}(s^{-1})+\Phi_{00}(s)Q_j(s^{-1})\right)^*\Pi_0Q_i(s^{-1})=Q_j(s^{-1})^{-1}Q_i(s^{-1}).
\end{split}
\end{equation*}
Therefore, by \eqref{FvsPhidef}, we obtain
\begin{equation}\label{f=QQ}
F_{ji}(s)=Q_j(s^{-1})^{-1}Q_i(s^{-1}),
\;\;\;\;\;\; i<j.
\end{equation}
In particular, the condition $Q_0=I$ gives
\begin{equation}\label{F10s}
F_{10}(s)=Q_1(s^{-1})^{-1}=\left[\left(\frac{1}{s}I-B_0\right)A_0^{-1}\right]^{-1}=sA_0(I-sB_0)^{-1}.
\end{equation}

\bex
Let $\Phi$ be the representation matrix of an OQW on $V=\{0,1,2\}$ of the form
$$\Phi=\begin{bmatrix}
      0 & \lceil C \rceil  &   \\
      \lceil A \rceil & 0 & \lceil C \rceil   \\
       & \lceil A \rceil & 0
    \end{bmatrix},\quad A=\frac{1}{2}\begin{bmatrix}
    -1 & 0 \\
    1 & \sqrt{2}
  \end{bmatrix},\quad
C=\frac{1}{2}\begin{bmatrix}
    1 & -\sqrt{2} \\
    -1 & 0
  \end{bmatrix}.
$$
Since $A^*A<I,$  the walk has an absorbing barrier in the frontier. Also, we have
$$(I-s\mathbb{Q}_1\Phi)=
\begin{bmatrix}
  I_4 & X & 0 \\
  0 & I_4 & 0 \\
  0 & Y & I_4
\end{bmatrix},\;
X=\frac{s}{4}
\begin{bmatrix}
  -1 & \sqrt{2} & \sqrt{2} & -2 \\
  1 & 0 & -\sqrt{2} & 0 \\
  1 & -\sqrt{2} & 0 & 0 \\
  -1 & 0 & 0 & 0
\end{bmatrix},\;
Y=\frac{s}{4}
\begin{bmatrix}
    -1 & 0 & 0 & 0 \\
    1 & \sqrt{2} & 0 & 0 \\
    1 & 0 & \sqrt{2} & 0 \\
    -1 & -\sqrt{2} & -\sqrt{2} & -2
  \end{bmatrix}.
$$
and
$$
F_{10}(s)=s\mathbb{P}_1\Phi(I-s\mathbb{Q}_1\Phi)^{-1}\mathbb{P}_0=
\frac{s}{4}
\begin{bmatrix}
  1 & 0 & 0& 0 \\
  -1 & -\sqrt{2} & 0 & 0 \\
  -1 & 0 & -\sqrt{2} & 0 \\
  1 & \sqrt{2} & \sqrt{2} & 2
\end{bmatrix}.
$$
The first two associated polynomials are given by
$$
Q_0(x)=I_4,\;\;Q_1(x):=2x
\begin{bmatrix}
  2 & 0 & 0 & 0 \\
  -\sqrt{2} & -\sqrt{2} & 0 & 0 \\
  -\sqrt{2} & 0 & -\sqrt{2} & 0 \\
  1 & 1 & 1 & 1
\end{bmatrix},
$$
from which we can calculate the product $Q_1(s^{-1})^{-1}Q_0(s^{-1})$, which equals $F_{10}(s)$ as expected. Then, for $\rho=\begin{bmatrix} a & b \\b^*& 1-a\end{bmatrix},$ we obtain
$$
p(\rho\otimes|0\rangle \to|1\rangle)=\lim_{s\uparrow 1}{\color{black}\mathrm{Tr}(F_{10}(s)\rho)}=\frac{1+\sqrt{2}\textmd{Re}(b)}{2}\in\left[\frac{2-\sqrt{2}}{4},\frac{2+\sqrt{2}}{4}\right],
$$
since $\textmd{Re}(b)\in[-1/2,1/2].$

\eex
\qee

\bex
Let $\gamma\in \mathbb{R}$ and $k_\gamma=2+2\gamma^2$ and $\Phi$ be the representation matrix of an OQW of the form
$$\Phi=\begin{bmatrix}
      \lceil B_0 \rceil & \lceil C_1 \rceil &  &  &  \\
      \lceil A_0 \rceil & \lceil B_1 \rceil &\lceil C_2 \rceil&  &  \\
       & \lceil A_1 \rceil & \lceil B_2 \rceil &\lceil C_3 \rceil &  \\
          &  & \ddots & \ddots &\ddots
    \end{bmatrix},\quad B_0=\frac{1}{\sqrt{k_\gamma}}\begin{bmatrix}
    -1 & \sqrt{2}\gamma \\
    0 & 1
  \end{bmatrix},\quad
A_0=\frac{1}{\sqrt{k_\gamma}}\begin{bmatrix}
    \sqrt{2}\gamma & 1 \\
    1 & 0
  \end{bmatrix}.
  $$
We notice that $F_{10}(s)$ does not depend on the blocks $A_k,B_k,C_k$ for $k=1,2,3,\ldots,$ thus such blocks can be chosen arbitrarily so that $A_k^*A_k+B_k^*B_k+C_k^*C_k=I$ for $k\geq 1$. Then, equation \eqref{firstVFdoqw2} gives
$$F_{10}(s)=\frac{s}{2+2\gamma^2-s}
\begin{bmatrix}
  2\gamma^2 & \frac{\sqrt{2}\gamma(2\gamma^2s+2-2\gamma^2-s)}{2+s+2\gamma^2} & \frac{\sqrt{2}\gamma(2\gamma^2s+2-2\gamma^2-s)}{2+s+2\gamma^2} & \frac{s+4\gamma^2s+4\gamma^4s+2+2\gamma^2}{2+s+2\gamma^2} \\
  \sqrt{2}\gamma & -\frac{2\gamma^2s}{2+s+2\gamma^2} & \frac{2\gamma^2+2-2\gamma^2s-s}{2+s+2\gamma^2} & \frac{\sqrt{2}\gamma(1+2\gamma^2)}{2+s+2\gamma^2} \\
  \sqrt{2}\gamma & \frac{2\gamma^2+2-2\gamma^2s-s}{2+s+2\gamma^2} & -\frac{2\gamma^2s}{2+s+2\gamma^2} & \frac{\sqrt{2}\gamma(1+2\gamma^2)}{2+s+2\gamma^2} \\
  1 & -\frac{\sqrt{2}\gamma s}{2+s+2\gamma^2} & -\frac{\sqrt{2}\gamma s}{2+s+2\gamma^2} & \frac{2\gamma^2s}{2+s+2\gamma^2}
\end{bmatrix},
$$
and, as expected, this is the same matrix obtained by formula \eqref{F10s}. For $\rho=\begin{bmatrix} a & b \\b^*& 1-a\end{bmatrix},$ we obtain, for every $\rho$, that
\begin{eqnarray}
\nonumber  p(\rho\otimes|0\rangle \to|1\rangle) &=& \lim_{s\uparrow 1}\mathrm{Tr}\left(F_{10}(s)\rho\right) \\
\nonumber   &=&\lim_{s\uparrow 1} \frac{4\gamma^4(as-a-s)+4\gamma\sqrt{2}(s-1)\textmd{Re}(b)(\gamma^2+1)+2\gamma^2(2as-3s-2a-1)-2-s}{(2+s+2\gamma^2)(-2+s-2\gamma^2)}=1.
\end{eqnarray}
We note that, in principle, we are able to obtain probabilities regarding vertices which are arbitrarily distant from one another but, as the distance between them increases, the task of performing explicit calculations may become unpractical. In such cases, it may be preferable to use the generating function (\ref{firstVFdoqw2}).
\eex
\qee

\section{An example of a QMC on a finite number of vertices}\label{sec5}

Let us first consider a walk induced by the block matrix on the $N+1$ nodes indexed as $\{0,1,\dots, N\}$,
\begin{equation*}\label{OpN}
\Phi=\begin{bmatrix} B & rI &  & & \\ tI & B & rI  & & \\  & tI & B & rI  & \\ & \ddots & \ddots & \ddots &  \\ &  & tI & B & rI\\  & &  & tI & B\end{bmatrix},\;\;\;0<r,t<1,
\end{equation*}
where if $B=\lceil \Phi_B\rceil$, $\Phi_B=V_1^*\cdot V_1+V_2^*\cdot V_2$, with
$$V_1=\sqrt{s}\begin{bmatrix} a & b \\ b & -a\end{bmatrix},\;\;\;V_2=\sqrt{s(1-a^2-b^2)}I_2.$$
We can write
$$B=s\begin{bmatrix} 1-b^2 & ab & ab & b^2 \\ ab & 1-2a^2-b^2 & b^2 & -ab \\ ab & b^2 & 1-2a^2-b^2 & -ab \\ b^2 & -ab & -ab & 1-b^2\end{bmatrix}.$$
For simplicity we assume $0<a,b,s<1$, $a^2+b^2<1$. In this way we have that $\mathrm{\mathrm{Tr}}(\Phi(X))=s\mathrm{\mathrm{Tr}}(X)$, so we suppose that $r+s+t=1$ in order to have that $\Phi$ is trace-preserving, with the exception of the first and last nodes (we remark that another restriction on $r,s,t$ will be needed, see below).

\medskip

By the classical symmetrization
$$\mathcal{R}=\mbox{diag}(R_0,R_1,\dots,R_{N}),\;\;\;R_i=\left(\sqrt{\frac{r}{t}}\right)^{i-1}I_4,\;\;\;i=1,\dots,N,\quad R_0=I_4,$$
we obtain
$$J=\mathcal{R}\Phi\mathcal{R}^{-1}=\begin{bmatrix} B & kI &  & & \\ kI & B & kI  & & \\  & kI & B & kI  & \\ & \ddots & \ddots & \ddots &  \\ &  & kI & B & kI\\  & &  & kI & B\end{bmatrix},\;\;\;k=\sqrt{rt}.
$$
The matrix-valued polynomials $\{Q_n\}_{n\geq0}$ defined by
\begin{align*}
Q_0(x)&=1,\quad Q_{-1}(x)=0,\\
xQ_0(x)&=Q_0(x)B+kQ_1(x),\\
xQ_i(x)&=kQ_{i-1}(x)+Q_i(x)B+kQ_{i+1}(x),\quad i=1,\dots,N-1,
\end{align*}
can be identified with the Chebyshev polynomials of the second kind $\{U_n\}_{n\geq0}$. Indeed, it is possible to see that $Q_n(x)=U_n\left((x-B)/2k\right),n\geq0$. Now, if we define
\begin{equation*}
R_{N+1}(x):=Q_N(x)(x-B)-kQ_{N-1}(x),
\end{equation*}
we have that the zeros of  det$(R_{N+1}(x))$ coincide with the eigenvalues of $J=\mathcal{R}\Phi\mathcal{R}^{-1}$. A simple calculation shows that
$$R_{N+1}(x)=kU_{N+1}\left(\frac{x-B}{2k}\right).$$
We would like to solve the equation det$(R_{N+1}(x))=0$. Recalling the representation
$$U_n\left(\frac{z}{2}\right)=\prod_{j=1}^n\left(z-2\cos\left(\frac{j\pi}{n+1}\right)\right),$$
we obtain, for the matrix-valued case at hand,
$$\mbox{det}(R_{N+1}(x))=k^4\mbox{det}\left(U_{N+1}\left(\frac{x-B}{2k}\right)\right)=k^4\mbox{det}\left[\prod_{j=1}^{N+1}\left(\frac{xI_4-B}{k}-2\cos\left(\frac{j\pi}{N+2}\right)I_4\right)\right]$$
$$=k^4\prod_{j=1}^{N+1} \mbox{det}\left[\left(\frac{xI_4-B}{k}-2\cos\left(\frac{j\pi}{N+2}\right)I_4\right)\right].$$
Noting that the eigenvalues of $B$ are $s$ and $s(1-2a^2-2b^2)$ (both with multiplicity 2) we have
\begin{align*}
&\mbox{det}\left[\left(\frac{xI_4-B}{k}-2\cos\left(\frac{j\pi}{N+2}\right)I_4\right)\right]\\
&=\mbox{det}\begin{bmatrix} \frac{x-s}{k}-2\cos\left(\frac{j\pi}{N+2}\right) & & & \\
0 & \frac{x-s}{k}-2\cos\left(\frac{j\pi}{N+2}\right) & & \\
 & & \frac{x-s(1-2a^2-2b^2)}{k}-2\cos\left(\frac{j\pi}{N+2}\right) & \\
  & & & \frac{x-s(1-2a^2-2b^2)}{k}-2\cos\left(\frac{j\pi}{N+2}\right)\end{bmatrix}\\
 &=\left[\frac{x-s}{k}-2\cos\left(\frac{j\pi}{N+2}\right)\right]^2\left[\frac{x-s(1-2a^2-2b^2)}{k}-2\cos\left(\frac{j\pi}{N+2}\right)\right]^2.
\end{align*}
Hence,
$$\mbox{det}(R_{N+1}(x))=k^4\prod_{j=1}^{N+1}\left[\frac{x-s}{k}-2\cos\left(\frac{j\pi}{N+2}\right)\right]^2\left[\frac{x-s(1-2a^2-2b^2)}{k}-2\cos\left(\frac{j\pi}{N+2}\right)\right]^2,\quad k=\sqrt{rt},$$
which is a polynomial of degree $4(N+1)$ having $2(N+1)$ distinct roots (all of multiplicity 2). Therefore, the roots are of the form
\begin{equation*}
x_{j}=s+2k\cos\left(\pi\frac{j+1}{N+2}\right),\;\;\;j=0,\dots,N,
\end{equation*}
\begin{equation*}
y_{j}=s(1-2a^2-2b^2)+2k\cos\left(\pi\frac{j+1}{N+2}\right),\;\;\;j=0,\dots,N,
\end{equation*}
all being of multiplicity 2, except in the case where the collection of zeros $x_N$ and $y_N$ overlap, so the multiplicity changes accordingly (see the example below). The expressions on the roots also make clear that we must have further restrictions on the values of $r,s$ and $t$ (recall $k=\sqrt{rt}$) so that $x_j, y_j\in [-1,1]$, for all $j=0,\dots,N$. For instance, by imposing $0<k<1/4$ we obtain a corresponding restriction on $s$ (we omit the details).

\medskip

The above root calculation should be compared with the classical case with a translation of $s$ units, for which the roots of $R_{N+1}$ are
$$x_{j}=s+2\sqrt{rt}\cos\left(\pi\frac{j+1}{N+1}\right),\;\;\;j=0,\dots,N,$$
once again regarding a random walk with a proper restriction on $r,s,t$ so that $x_j\in[-1,1]$, for all $j$.

\medskip

Now we compute the matrix weights on the zeros above. Such calculation needs to take in consideration the fact that each root is double (we omit the discussion for the case of larger multiplicities). In this case the residue calculation gives us that
\beq\label{doublef} W_j=g_j'(\lambda_j),\;\;\;g_j(\lambda):=-(\lambda_j-\lambda)^2(J-\lambda I)_{00}^{-1},\;\;\;\lambda_j=x_{j}, y_{j},\;\;\;j=0,\dots,N,\eeq
an expression which can be deduced from (see \cite{grunb})
$$
(J-\lambda I)_{ij}^{-1}=\sum_{k=0}^N \frac{P_i^*(\lambda_k)W_k P_j(\lambda_k)}{\lambda_k-\lambda},
$$
and noting that this corresponds to the Laurent sum of the operator on the left-hand side except for the sign change $\lambda_k-\lambda=-(\lambda-\lambda_k)$. With formula \eqref{doublef}, a calculation shows that for every $N$ we have a corresponding set of multiples of the matrices given by
$$W_{a,b;1}:=\frac{1}{2(a^2+b^2)}\begin{bmatrix}
2a^2+b^2 & ab& ab & b^2 \\
ab & b^2 & b^2 & -ab \\
ab & b^2 & b^2 & -ab \\
b^2 & -ab & -ab & 2a^2+b^2\end{bmatrix},\;\;\;
W_{a,b;2}:=\frac{1}{2(a^2+b^2)}\begin{bmatrix}
b^2 & -ab& -ab & -b^2 \\
-ab & b^2+2a^2 & -b^2 & ab \\
-ab & -b^2 & b^2+2a^2 & ab \\
-b^2 & ab & ab & b^2\end{bmatrix}.
$$
More precisely, we have a collection of $4(N+1)$ roots with weights
$$\psi(x_{j})=\frac{2}{N+2}\sin^2\left(\pi\frac{j+1}{N+2}\right)W_{a,b;1},\;\;\;j=0,\dots,N,$$
$$\psi(y_{j})=\frac{2}{N+2}\sin^2\left(\pi\frac{j+1}{N+2}\right)W_{a,b;2},\;\;\;j=0,\dots,N.
$$
This should be compared with the classical setting, recalling that in such case,
\begin{equation}\label{clsss}
\psi(x_{j})=\frac{2}{N+2}\sin^2\left(\pi\frac{j+1}{N+2}\right)=\frac{1}{2pq(N+2)}(4pq-x_{j}^2),\;\;\;j=0,\dots,N.
\end{equation}
We note a few basic properties of $W_{a,b;1}$ and $W_{a,b;2}$. First, both are positive semidefinite matrices with eigenvalues $0$ and $1$ (multiplicity 2). Moreover, seen as linear maps, $W_{a,b;1}$ is trace-preserving, whereas $W_{a,b;2}$ transforms densities into traceless matrices. Also $W_{a,b;1}$ admits the following Kraus representation
$$W_{a,b;1}=\sum_{i=1}^3 W_i^1\otimes\ov W_i^1,\;\;\;W_1^1=\frac{1}{2(a^2+b^2)}\begin{bmatrix} a & b \\ b & -a\end{bmatrix},\;\;\;W_2^1=\frac{a}{2(a^2+b^2)}I_2,\;\;\;W_3^1=\frac{b}{2(a^2+b^2)}I_2,$$
from which we conclude that such weight represents a completely positive map. However, $W_{a,b;2}$ does not represent a positive map in general, as illustrated by an inspection with certain density examples.

\medskip

For a specific instance of the above take $N=4$ (5 sites), so we have 20 roots, with weights
$$\frac{1}{3}W_{a,b;1},\;\;\;\frac{1}{3}W_{a,b;2},$$
associated with zeros $s$ and $s(1-2a^2-2b^2)$ respectively; weights
$$\frac{1}{4}W_{a,b;1},\;\;\;\frac{1}{4}W_{a,b;2},$$
associated with zeros $s\pm k$, $s(1-2a^2-2b^2)\pm k$ respectively; and weights
$$\frac{1}{12}W_{a,b;1},\;\;\;\frac{1}{12}W_{a,b;2},$$
associated with zeros $s\pm \sqrt{3}k$, and $s(1-2a^2-2b^2)\pm \sqrt{3}k$ respectively. If, moreover, $s=a=b=k=1/2$, we have
$$\{x_{j}\}_{j=0\dots 4}=\left\{-\frac{\sqrt{3}}{2},-\frac{1}{2},0,\frac{1}{2},\frac{\sqrt{3}}{2}\right\},\;\;\;\{y_{j}\}_{j=0\dots 4}=\left\{\frac{-\sqrt{3}+1}{2},0,\frac{1}{2},1,\frac{\sqrt{3}+1}{2}\right\},$$
each with multiplicity 2 except for $0$ and $1/2$ with multiplicity $4$ (noting that in this case, $1-2a^2-2b^2=0$). This should be compared with the classical setting, see \eqref{clsss}.

\section{An example of a QMC on $\mathbb{Z}_{\geq0}$}\label{sec6}

Consider the walk induced by the block matrix on $\mathbb{Z}_{\geq0}$ given by
\beq\label{comp1ex}
\Phi=\begin{bmatrix} 0 & C & & & 0 \\
                    A & 0 & C &  & & \\
                     & A & 0 & C & \\
                      0& & \ddots & \ddots & \ddots\end{bmatrix},
\eeq
where $A$ and $C$ are the compact forms (see \eqref{compform}) of $R_1 \otimes \ov{R_1}+R_2 \otimes \ov{R_2}$ and $L_1 \otimes \ov{L_1}+L_2 \otimes \ov{L_2}$, respectively, and
$$
L_1=\sqrt{p/2}I_2,\quad L_2=\sqrt{(1-p)/2}\begin{bmatrix} 0 & 1 \\ 1 & 0\end{bmatrix},\quad R_1=\sqrt{q/2}\begin{bmatrix} 1 & 0 \\ 0 & -1\end{bmatrix},\quad R_2=\sqrt{(1-q)/2}\begin{bmatrix} 0 & 1 \\ 1 & 0\end{bmatrix}.
$$
Observe that $R_1^*R_1+R_2^*R_2+L_1^*L_1+L_2^*L_2=I_2$. Therefore,
$$
A=\frac{1}{2}\begin{bmatrix} q & 0 & 1-q\\ 0 & 1-2q &0\\ 1-q & 0& q\end{bmatrix},\quad C=\frac{1}{2}\begin{bmatrix} p & 0 & 1-p\\ 0 & 1 &0\\ 1-p & 0& p\end{bmatrix}.
$$
The matrices $A$ and $B$ are simultaneously diagonalizable, i.e.,
\beq\label{ACU}
A=\mathcal{U}\begin{bmatrix} 1/2 &  & \\  & 1/2-q &\\  & & q-1/2\end{bmatrix}\mathcal{U}^*,\quad C=\mathcal{U}\begin{bmatrix}1/2 &  & \\  & 1/2 &\\  & & p-1/2\end{bmatrix}\mathcal{U}^*,\quad \mathcal{U}=\frac{1}{\sqrt{2}}\begin{bmatrix} 1 & 0 & -1\\ 0 & \sqrt{2} &0\\ 1 & 0& 1\end{bmatrix}.
\eeq
Choosing
$$
\Pi_n=\begin{bmatrix} 1 &  & \\  & (1-2q)^n &\\  & & \left(\displaystyle\frac{1-2q}{1-2p}\right)^n\end{bmatrix},
$$
we can symmetrize the operator \eqref{comp1ex}, getting that each of the nonzero blocks are given by
$$
\frac{1}{2}\mathcal{U}\begin{bmatrix} 1 &  & \\  & \sqrt{1-2q} &\\  & & \sqrt{(1-2p)(1-2q)}\end{bmatrix}\mathcal{U}^*.
$$
The Stieltjes transform associated with \eqref{comp1ex} is given by
\begin{equation}\label{Sti1ex}
B(z;W)=2\mathcal{U}\begin{bmatrix} z-\sqrt{z^2-1} &  & \\  & \displaystyle \frac{z-\sqrt{z^2-(1-2q)}}{1-2q} &\\  & &\displaystyle \frac{z-\sqrt{z^2-(1-2p)(1-2q)}}{(1-2p)(1-2q)}\end{bmatrix}\mathcal{U}^*.
\end{equation}
Therefore, we get an absolutely continuous weight matrix given by
$$
dW(x)=\frac{2}{\pi}\mathcal{U}D(x)\mathcal{U}^*dx,
$$
where
$$
D(x)=\begin{bmatrix}
\left[\omega_1(x)\right]_+&&\\
&\left[\omega_2(x)\right]_+&\\
&&\left[\omega_3(x)\right]_+
\end{bmatrix},
$$
where
\begin{equation}\label{w1w2w3}
\omega_1(x)=\sqrt{1-x^2},\quad\omega_2(x)=\frac{\sqrt{1-2q-x^2}}{1-2q},\quad\omega_3(x)=\frac{\sqrt{(1-2p)(1-2q)-x^2}}{(1-2p)(1-2q)}.
\end{equation}
Here we are using the notation $\left[f(x)\right]_+=f(x)$ if $f(x)\geq0$ and 0 otherwise. Similar results can be obtained if we do not consider the compact form.

\medskip

Now consider the same walk as before in \eqref{comp1ex}, but adding a matrix $B$ at the upper-left corner, i.e.
\beq\label{comp1ex2}
\widetilde{\Phi}=\begin{bmatrix} B & C & & & 0 \\
                    A & 0 & C &  & & \\
                     & A & 0 & C & \\
                      0& & \ddots & \ddots & \ddots\end{bmatrix},
\eeq
where $B$ is a matrix which we assume it can be written as
\beq\label{bbb}
B=\frac{1}{2}\mathcal{U}\begin{bmatrix}
b_1&&\\
&b_2&\\
&&b_3
\end{bmatrix}\mathcal{U}^*,
\eeq
with $\mathcal{U}$ defined by \eqref{ACU}. According to Theorem 2.6 of \cite{dette}, the Stieltjes transform $B(z;\widetilde W)$ associated with \eqref{comp1ex2} can be written as $B(z;\widetilde W)=(B(z;W)^{-1}-B)^{-1}$. Since we are assuming \eqref{bbb} and taking in mind \eqref{Sti1ex}, we obtain
\begin{equation*}
B(z;\widetilde W)=2\mathcal{U}\begin{bmatrix} \displaystyle\frac{1}{z-\sqrt{z^2-1}}-b_1 &  & \\  & \displaystyle \frac{1-2q}{z-\sqrt{z^2-(1-2q)}}-b_2 &\\  & &\displaystyle \frac{(1-2p)(1-2q)}{z-\sqrt{z^2-(1-2p)(1-2q)}}-b_3\end{bmatrix}^{-1}\mathcal{U}^*.
\end{equation*}
After rationalization and some computations we obtain
\begin{equation}\label{Sti1ex2}
B(z;\widetilde W)=2\mathcal{U}\begin{bmatrix} \displaystyle\frac{-z+b_1+\sqrt{z^2-1}}{2b_1z-1-b_1^2} &  & \\  & \displaystyle \frac{-z+b_2-\sqrt{z^2-(1-2q)}}{2b_2z-1+2q-b_2^2} &\\  & &\displaystyle \frac{-z+b_3+\sqrt{z^2-(1-2p)(1-2q)}}{2b_3z-(1-2p)(1-2q)-b_3^2}\end{bmatrix}\mathcal{U}^*.
\end{equation}
Therefore the weight matrix is given by $\widetilde{W}=\widetilde{W}_{ac}+\widetilde{W}_{d}$, where the absolutely continuous part is given by
$$
d\widetilde{W}_{ac}(x)=\frac{2}{\pi}\mathcal{U}\begin{bmatrix}
\displaystyle\frac{\left[\sqrt{1-x^2}\right]_+}{1+b_1^2-2b_1x}&&\\
&\displaystyle\frac{\left[\sqrt{1-2q-x^2}\right]_+}{1-2q+b_2^2-2b_2x}&\\
&&\displaystyle\frac{\left[\sqrt{(1-2p)(1-2q)-x^2}\right]_+}{(1-2p)(1-2q)+b_3^2-2b_3x}
\end{bmatrix}\mathcal{U}^*dx.
$$
Observe that the denominators are always nonnegative in the range of the definition of each square root. The discrete part $\widetilde{W}_{d}$ is given by three Dirac deltas located at the poles of the Stieltjes transform \eqref{Sti1ex2}, i.e.
$$
\widetilde{W}_{d}(x)=\mathcal{U}\begin{bmatrix}
\widetilde{W}\left(\{z_1\}\right)\delta_{z_1}(x)&&\\
&\widetilde{W}\left(\{z_2\}\right)\delta_{z_2}(x)&\\
&&\widetilde{W}\left(\{z_3\}\right)\delta_{z_3}(x)
\end{bmatrix}\mathcal{U}^*,
$$
where
$$
z_1=\frac{1+b_1^2}{2b_1},\quad z_2=\frac{1-2q+b_2^2}{2b_2},\quad z_3=\frac{(1-2p)(1-2q)+b_3^2}{2b_3},
$$
and
\begin{align*}
\widetilde{W}\left(\{z_1\}\right)&=\frac{b_1^2-1}{b_1^2}\mathbf{1}_{\{b_1^2>1\}},\\
\widetilde{W}\left(\{z_2\}\right)&=\frac{b_2^2-(1-2q)}{b_2^2}\mathbf{1}_{\{b_2^2>1-2q\}},\\
\widetilde{W}\left(\{z_3\}\right)&=\frac{b_3^2-(1-2p)(1-2q)}{b_3^2}\mathbf{1}_{\{b_3^2>(1-2p)(1-2q)\}}.
\end{align*}
Observe that in principle $b_1,b_2$ and $b_3$ can be taken as any real numbers, but we are interested in finding under what conditions the points $z_1,z_2$ and $z_3$ are located inside the interval $[-1,1]$ (so that all the support of $\widetilde{W}$ is inside the interval $[-1,1]$). By the definition it is possible to see that $|z_1|\leq1,|z_2|\leq1,|z_3|\leq1,$ if and only if  $b_1=1$, and
\begin{align*}
b_2&\in[-1-\sqrt{2q},-1+\sqrt{2q}]\cup[1-\sqrt{2q},1+\sqrt{2q}],\\
b_3&\in[-1-\sqrt{2(p+q-2pq)},-1+\sqrt{2(p+q-2pq)}]\cup[1-\sqrt{2(p+q-2pq)},1+\sqrt{2(p+q-2pq)}].
\end{align*}
Joining this with the conditions under we have positive jumps, we have that $\widetilde{W}\left(\{z_1\}\right)=0$ and $\widetilde{W}\left(\{z_2\}\right), \widetilde{W}\left(\{z_3\}\right)$ are positive if
\begin{align*}
b_2&\in[-1-\sqrt{2q},-\sqrt{1-2q})\cup(\sqrt{1-2q},1+\sqrt{2q}],\\
b_3&\in[-1-\sqrt{2(p+q-2pq)},-\sqrt{(1-2p)(1-2q)})\cup(\sqrt{(1-2p)(1-2q)},1+\sqrt{2(p+q-2pq)}].
\end{align*}

The particular case where $B=A$ is given by $b_1=1, b_2=1-2q,b_3=2q-1$. Therefore $z_1=1,z_2=1-q, z_3=p+q-1$, $\widetilde{W}\left(\{z_1\}\right)=\widetilde{W}\left(\{z_2\}\right)=0$ and
$$
\widetilde{W}\left(\{z_3\}\right)=\frac{2(p-q)}{1-2q}\mathbf{1}_{\{p>q\}}.
$$
The weight matrix is then given by $\widetilde{W}=\widetilde{W}_{ac}+\widetilde{W}_{d}$, where
\begin{equation}\label{ABeqc}
d\widetilde{W}_{ac}(x)=\frac{1}{\pi}\mathcal{U}\begin{bmatrix}
\left[\displaystyle\sqrt{\frac{1+x}{1-x}}\right]_+&&\\
&\displaystyle\frac{\left[\sqrt{1-2q-x^2}\right]_+}{(1-2q)(1-q-x)}&\\
&&\displaystyle\frac{\left[\sqrt{(1-2p)(1-2q)-x^2}\right]_+}{(1-2q)(1-p-q+x)}
\end{bmatrix}\mathcal{U}^*dx.
\end{equation}
and
\begin{equation}\label{ABeqd}
\widetilde{W}_{d}(x)=\frac{p-q}{1-2q}\mathbf{1}_{\{p>q\}}\begin{bmatrix}
1&0&-1\\
0&0&0\\
-1&0&1
\end{bmatrix}\delta_{p+q-1}(x).
\end{equation}
Observe that in this situation, as expected, the support of $\widetilde{W}$ is inside the interval $[-1,1]$.

\medskip

Let us now study recurrence of this QMC in terms of the corresponding weight matrices. Note that the QMC determined by  (\ref{comp1ex}) is such that vertex $0$ admits a transition to an absorbing state, so we have the transience of this walk with respect to such site. Let us prove this in terms of the associated measure. First, recall that the trace is invariant by the change of coordinates $\mathcal{U}$ which, on its turn, does not depend on $x$. Therefore, we need only to examine the behavior of $\omega_1$ and $\omega_3$ in \eqref{w1w2w3}. Regarding $\omega_1$, a calculation gives that
$$
\lim_{z\uparrow 1}\int_{-1}^1 \frac{\sqrt{1-x^2}}{1-zx}dx=\lim_{z\uparrow 1}\frac{\pi(z^2-1+\sqrt{1-z^2})}{z^2\sqrt{1-z^2}}=\pi,
$$
so the above limit is finite. Regarding $\omega_3$, note that since $0<p,q<1$, we have $a:=(1-2p)(1-2q)>0$ if and only if both $p$ and $q$ are greater than $1/2$ or both are less than $1/2$. If this is the case, we have that $\omega_3(x)\geq 0$ if $x\in(-\sqrt{a},\sqrt{a})$. If we write $q=p+\epsilon$ (with $\epsilon\in (\frac{1}{2}-p,1-p)$ if $\frac{1}{2}<p<1$), we obtain
\beq
\label{integral1}\lim_{z\uparrow 1}\int_{-\sqrt{a}}^{\sqrt{a}} \frac{\sqrt{a-x^2}}{1-zx}dx=\pi(1-\sqrt{4p(1-p)+2\epsilon(1-2p)}),
\eeq
which is also a finite number (as expected, the term inside the root is always positive under the above restrictions). A similar reasoning holds in the case $0<p<\frac{1}{2}$, where we write $q=p+\epsilon$, with $\epsilon\in (-p,\frac{1}{2}-p)$. In the case that $\omega_3$ does not have a positive part, the trace computation is determined by $\omega_1$. Since $\mathcal{U}^*\rho$ is also a density matrix we conclude that,  in every case, site $0$ is transient with respect to any initial density.

\medskip

Now considering (\ref{comp1ex2}) with $B=A$ (see \eqref{ABeqc} and \eqref{ABeqd}), we have, regarding $\widetilde\omega_1$, that
$$
\lim_{z\uparrow 1}\int_{-1}^1 \frac{1}{1-zx}\sqrt{\frac{1+x}{1-x}}dx=\lim_{z\uparrow 1}\frac{\pi(1+z-\sqrt{1-z^2})}{z\sqrt{1-z^2}}=\infty.
$$
Regardind $\widetilde\omega_3$, we note that the denominator is positive if $x\in(-\sqrt{a},\sqrt{a})$, which can be seen as in the transient walk above (i.e., consider the cases for which $p,q\in (0,\frac{1}{2})$ or $p,q\in (\frac{1}{2},1)$). But then the limit to be examined is the same as for the transient walk, namely, eq. \eqref{integral1}, which is finite. We have concluded that recurrence of site $0$ depends on the initial choice of density matrix. For instance, the densities
$$\rho_\alpha=\begin{bmatrix} 1 & 0 \\ 0 & 0\end{bmatrix}\otimes|0\rangle\langle 0|,\;\;\;\rho_\beta=\begin{bmatrix} 0 & 0 \\ 0 & 1\end{bmatrix}\otimes|0\rangle\langle 0|,$$
are such that site $0$ is recurrent with respect to $\rho_\alpha$  but transient with respect to $\rho_\beta$. More generally, site $0$ will be recurrent with respect to any density matrix $\rho\otimes |0\rangle\langle 0|$ for which $\rho_{11}>0$. It would be interesting to find examples of matrices $B$ at the block position $(0,0)$ for which the resulting walks are irreducible (if this is in fact possible, a guess would be to obtain a change of coordinates $\mathcal{V}$ distinct from $\mathcal{U}$).

\begin{remark}
If $B$ in \eqref{bbb} is not simultaneously diagonalizable with $A$ and $C$, it is possible to derive again the weight matrix assuming that $B=\frac{1}{2}\mathcal{V}\mbox{diag}\{b_1,b_2,b_3\}\mathcal{V}^*$, where $\mathcal{V}$ is unitary. The corresponding weight matrix will be also unitarily diagonalizable.
\end{remark}

\section{QMCs on $\mathbb{Z}$}\label{sec7}

In this section, we treat the case of tridiagonal QMCs on the real line, that is, the set of vertices $V$ will consist of the integers, thus the walk will have one-step transition probabilities from $\ket{i}$ to $\ket{i-1},\ket{i}$ or $\ket{i+1}$ and there are no barriers. Starting from a tridiagonal QMC $\Phi$ on $\mathbb{Z}$, where each of the blocks of the matrix representation is of order $N^2\times N^2$, we will construct a new tridiagonal QMC on $\mathbb{Z}_{\geq0}\times\{1,2\}$, where each of the blocks of the matrix representation is of dimension $2N^2\times 2N^2$ with a possible barrier on site $\ket{0}$. This is what we call the \textbf{folding trick} and was introduced for the first time in \cite{Ber}. Finally, recurrence of this type of walks will be discussed via an application of the Stieltjes transform.

\medskip

Consider then the matrix representation for a tridiagonal QMC on $\mathbb{Z}$, given by
\begin{equation}\label{PhiinZ}
\Phi=\left[\begin{array}{ccc|ccccc}
\ddots & \ddots &  &  &  &  &  &  \\
  \ddots & B_{-2} & C_{-1} &  &  &  &  &  \\
   & A_{-2} &B_{-1} & C_{0} &  &  &  &  \\
  \hline
   &    & A_{-1} & B_{0} & C_{1} & & &  \\
    &  &    & A_{0} & B_{1} & C_{2} & & \\
    & & &    & A_{1} & B_{2} & C_{3} &  \\
     &  &  &  &  & \ddots & \ddots &\ddots
\end{array}\right],
\end{equation}
where each block $A_k,B_k,C_k$ is an $N^2\times N^2$ matrix given by a summation
$$X_k=\sum_{r=1}^{t_k}\lceil Y_{r}\rceil,\;\;Y_{r}\in M_N(\mathbb{C}),\;\;\lceil Y_r\rceil=Y_r\otimes \overline{Y_r},$$
and we assume that there exists a sequence of Hermitian matrices $(E_n)_{n\in\mathbb{Z}}\in M_{N^2}(\mathbb{C})$ and non-singular matrices $(R_n)_{n\in\mathbb{Z}}\in M_{N^2}(\mathbb{C})$ such that
\begin{equation}\label{condiequiv}
\begin{split}
  A_n^*R_{n+1}^*R_{n+1} &= R_{n}^*R_{n}C_{n+1},  \; n\geq0\\
  R_{-n-1}^*R_{-n-1}C_{-n} &=  A_{-n-1}^*R_{-n}^*R_{-n}, \; n\geq 0,
\end{split}
\qquad R_nB_n=E_nR_n,\; n\in\mathbb{Z}.
\end{equation}
The previous conditions coincide with those of Theorem \ref{CondiDette} when we consider the first line with the walk restricted to $\mathbb{Z}_{\geq0}$ and the second line with the walk restricted to $\mathbb{Z}_{<0}.$ Let us define
$$
\Pi_j:=R_j^*R_j\in M_{N^2}(\mathbb{C}),\;\;j\in\mathbb{Z}.
$$
Consider the two independent families of matrix-valued polynomials defined recursively from \eqref{PhiinZ} as
\begin{equation}\label{polsZgeneralOQW}
\begin{split}
Q_0^1(x) &= I_{N^2}, \quad Q_0^2(x)=0,\\
Q_{-1}^1(x) &= 0,\quad Q_{-1}^2(x)=I_{N^2}, \\
xQ_n^\alpha (x) &= Q_{n+1}^\alpha(x)A_n+Q_{n}^\alpha(x)B_n+Q_{n-1}^\alpha(x)C_n,\quad \alpha=1,2,\;\;n\in\mathbb{Z}.
\end{split}
\end{equation}
and the block vectors $Q^\alpha(x)=\left(\ldots,Q_{-2}^\alpha(x),Q_{-1}^\alpha(x),Q_{0}^\alpha(x),Q_{1}^\alpha(x),Q_{2}^\alpha(x),\ldots\right),$ $\alpha=1,2,$ are linearly independent solutions, depending on the initial values at $n=0$, of the eigenvalue equation $xQ^\alpha(x)=Q^\alpha(x)\Phi.$

\begin{figure}[h!]
$$\begin{tikzpicture}[>=stealth',shorten >=1pt,auto,node distance=2cm]
\node[state] (q-2)      {$-2$};
\node[state]         (q-1) [right of=q-2]  {$-1$};
\node[state]         (q0) [right of=q-1]  {$0$};
\node[state]         (q1) [right of=q0]  {$1$};
\node[state]         (q2) [right of=q1]  {$2$};
\node         (qr) [right of=q2]  {$\ldots$};
\node         (ql) [left of=q-2]  {$\ldots$};
\path[->]          (ql)  edge         [bend left=15]  node[auto] {$A_{-3}$}     (q-2);
\path[->]          (q-2)  edge         [bend left=15]   node[auto] {$A_{-2}$}     (q-1);
\path[->]          (q-1)  edge         [bend left=15]   node[auto] {$A_{-1}$}     (q0);
\path[->]          (q0)  edge         [bend left=15]   node[auto] {$A_0$}     (q1);
\path[->]          (q1)  edge         [bend left=15]   node[auto] {$A_1$}     (q2);
\path[->]          (q2)  edge         [bend left=15]   node[auto] {$A_2$}     (qr);
\path[->]          (qr)  edge         [bend left=15] node[auto] {$C_{2}$}       (q2);
\path[->]          (q2)  edge         [bend left=15]   node[auto] {$C_{1}$}     (q1);
\path[->]          (q1)  edge         [bend left=15]  node[auto] {$C_{1}$}      (q0);
\path[->]          (q0)  edge         [bend left=15]   node[auto] {$C_{0}$}     (q-1);
\path[->]          (q-1)  edge         [bend left=15]  node[auto] {$C_{-1}$}      (q-2);
\path[->]          (q-2)  edge         [bend left=15]  node[auto] {$C_{-2}$}      (ql);
  \draw [->] (q1) to[loop above]node[auto] {$B_{1}$}  (q1);
   \draw [->] (q2) to[loop above]node[auto] {$B_{2}$}  (q2);
    \draw [->] (q-2) to[loop above] node[auto] {$B_{-2}$} (q-2);
     \draw [->] (q-1) to[loop above]node[auto] {$B_{-1}$}  (q-1);
      \draw [->] (q0) to[loop above]node[auto] {$B_{0}$}  (q0);
\end{tikzpicture}$$
\caption{QMC $\Phi$ on $\mathbb{Z}$.}
\label{originalQMConZ}
\end{figure}
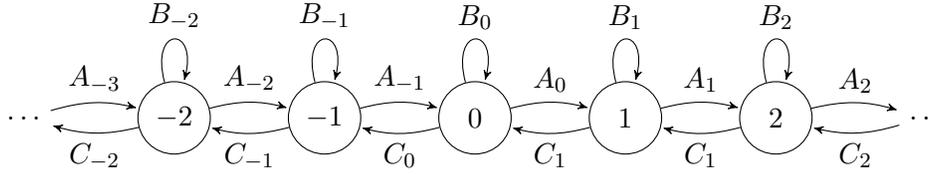

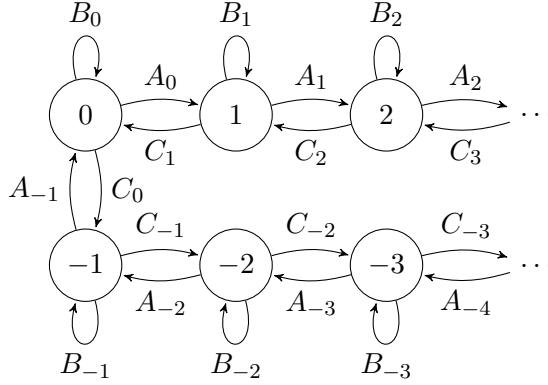
\begin{figure}[h]
$$\begin{tikzpicture}[>=stealth',shorten >=1pt,auto,node distance=2cm]
  \node[state] (q1)      {$0$};
  \node[state] (q2) [right of=q1]  {$1$};
  \node[state] (q3) [right of=q2]  {$2$};
  \node         (q4) [right of=q3]  {$\ldots$};
  \node[state] (q-1) [below of=q1]     {$-1$};
  \node[state] (q-2) [right of=q-1]  {$-2$};
  \node[state] (q-3) [right of=q-2]  {$-3$};
  \node         (q-4) [right of=q-3]  {$\ldots$};
  \path[->]          (q1)  edge         [bend left=15]   node[auto] {$A_0$}     (q2);
  \path[->]          (q2)  edge         [bend left=15]   node[auto] {$A_1$}     (q3);
  \path[->]          (q3)  edge         [bend left=15]   node[auto] {$A_2$}     (q4);
  \path[->]          (q2)  edge         [bend left=15]   node[auto] {$C_1$}     (q1);
  \path[->]          (q3)  edge         [bend left=15]   node[auto] {$C_2$}     (q2);
  \path[->]          (q4)  edge         [bend left=15]   node[auto] {$C_3$}     (q3);
  \draw [->]         (q1) to[loop above]  node[auto] {$B_0$} (q1);
  \draw [->]         (q2) to[loop above]  node[auto] {$B_1$}(q2);
  \draw [->]         (q3) to[loop above]  node[auto] {$B_2$}(q3);
  \path[->]          (q-1)  edge         [bend left=15]   node[auto] {$C_{-1}$}     (q-2);
  \path[->]          (q-2)  edge         [bend left=15]   node[auto] {$C_{-2}$}     (q-3);
  \path[->]          (q-3)  edge         [bend left=15]   node[auto] {$C_{-3}$}     (q-4);
  \path[->]          (q-2)  edge         [bend left=15]   node[auto] {$A_{-2}$}     (q-1);
  \path[->]          (q-3)  edge         [bend left=15]   node[auto] {$A_{-3}$}     (q-2);
  \path[->]          (q-4)  edge         [bend left=15]   node[auto] {$A_{-4}$}     (q-3);
    \draw [->]       (q-1) to[loop below]  node[auto] {$B_{-1}$} (q-1);
  \draw [->]         (q-2) to[loop below]  node[auto] {$B_{-2}$}(q-2);
  \draw [->]         (q-3) to[loop below]  node[auto] {$B_{-3}$}(q-3);
  \path[->]          (q1)  edge         [bend left=15]   node[auto] {$C_0$}     (q-1);
  \path[->]          (q-1)  edge         [bend left=15]   node[auto] {$A_{-1}$}     (q1);
\end{tikzpicture}$$
\caption{Folded walk of $\Phi$ on $\mathbb{Z}_{\geq0}\times\{1,2\}$ given by $\breve{\Phi}.$}
\label{foldingQMC}
\end{figure}

As in the classical case, we introduce the block tridiagonal matrix
\begin{equation*}\label{Phifold}
\breve{\Phi}=
\begin{bmatrix}
  G_0 & N_1 &  & & \\
  M_0&G_1 & N_2   &&  \\
  &M_1&G_2& N_3    & \\
   &  & \ddots &\ddots &\ddots
\end{bmatrix},
\end{equation*}
where each block entry is a $2N^2\times 2N^2$ matrix, given by
$$
\begin{array}{rlrll}
  G_0=&\begin{bmatrix}
        B_0 & A_{-1}\\
        C_0 & B_{-1}
      \end{bmatrix}, &
  M_n=&\begin{bmatrix}
        A_{n} & 0 \\
        0 & C_{-n-1}
      \end{bmatrix},&n\geq 0, \\
   G_n=&\begin{bmatrix}
        B_{n} & 0 \\
        0 & B_{-n-1}
      \end{bmatrix}, &
   N_n=&\begin{bmatrix}
        C_{n} & 0 \\
        0 & A_{-n-1}
      \end{bmatrix},&n\geq 1.
\end{array}
$$
The term \emph{folding trick} comes from the transformation of the original walk $\Phi$, whose graph is represented in Figure \ref{originalQMConZ},
to the QMC described by $\breve{\Phi},$ which is represented by the folded walk in Figure \ref{foldingQMC}.

Note that $\breve{\Phi}$ is a block tridiagonal matrix on $\mathbb{Z}_{\geq0},$ thereby we can apply all the properties we have seen in previous sections. The following polynomials are defined in terms of \eqref{polsZgeneralOQW},
\begin{equation}\label{blobloQn}
\mathcal{Q}_n(x)=
\begin{bmatrix}
  Q_{n}^1(x) &Q_{-n-1}^1(x) \\
    Q_{n}^2(x)& Q_{-n-1}^2(x)
\end{bmatrix}\in M_{2N^2}(\mathbb{C}),\;\;n\geq 0,
\end{equation}
and these satisfy
\begin{align*}
x\mathcal{Q}_{0}(x) =& \mathcal{Q}_{1}(x)M_0+\mathcal{Q}_{0}(x)G_0,\;\; \mathcal{Q}_0(x)=I_{2N^2}, \\
x\mathcal{Q}_{n}(x) =& \mathcal{Q}_{n+1}(x)M_n+\mathcal{Q}_{n}(x)G_n+\mathcal{Q}_{n-1}(x)N_n,\;\;n=1,2,\ldots
\end{align*}
The leading coefficient of $\mathcal{Q}_{n}(x)$ is always a nonsingular matrix. Moreover, for
$$
\breve{R}_n:=\begin{bmatrix}
       R_n & 0_{N^2} \\
       0_{N^2} & R_{-n-1}
     \end{bmatrix},\;n\geq 0,\quad
\breve{E}_0:=\begin{bmatrix}
               E_0 & R_0A_{-1}R_{-1}^{-1} \\
               R_{-1}C_0R_0^{-1} & E_{-1}
             \end{bmatrix},\quad
\breve{E}_n:=\begin{bmatrix}
               E_n & 0_{N^2} \\
               0_{N^2} & E_{-n-1}
             \end{bmatrix},\;n\geq 1,
$$
we see that the block matrices of $\breve{\Phi}$ satisfy the conditions \eqref{condiequiv} for $n\geq 0:$
\begin{equation*}\label{condiequivbreve}
M_{n}^*\breve{R}_{n+1}^*\breve{R}_{n+1}  =  \breve{R}_{n}^*\breve{R}_{n}N_{n+1}, \;\;\;   \breve{R}_nG_n=\breve{E}_n\breve{R}_n,
\end{equation*}
where matrices $\breve{R}_{n}$ are non-singular and $\breve{E}_n$ are Hermitian for all $n\geq 0$. Defining
$$
\breve{\Pi}_j:=\breve{R}_j^*\breve{R}_j\in M_{2N^2}(\mathbb{C}),\;\;j=0,1,2,\ldots,
$$
the correspondence between $\breve{\Pi}_j$ and $\Pi_j$ is
$$
\breve{\Pi}_j:=\begin{bmatrix}
                 \Pi_j & 0_{N^2} \\
                 0_{N^2} & \Pi_{-j-1}
               \end{bmatrix}.
$$
By \cite{dette} (see also \eqref{Kmc44}), there exists a weight matrix $W$ leading to the Karlin-McGregor formula for $\breve{\Phi}:$
\begin{equation}\label{KMFqmc}
\breve{\Phi}_{ji}^{(n)}=\breve{\Pi}_j\int_\mathbb{R} x^n\mathcal{Q}_{j}^*(x)dW(x)\mathcal{Q}_{i}(x).
\end{equation}
Once we have found the weight matrix appearing on \eqref{KMFqmc}, we can also obtain the blocks $\Phi_{ji}^{(n)}$ of the original walk $\Phi.$ The key for this operation is the following proposition:

\begin{pro}\label{Phibvsn} Assume that $\Phi$ is a QMC of the form \eqref{PhiinZ}.
The relation between $\breve{\Phi}_{ij}^{(n)}$ and $\Phi_{ij}^{(n)}$ is
\begin{equation}\label{PHIbrevevshat}
\breve{\Phi}_{ji}^{(n)}=\begin{bmatrix}
               \Phi_{ji}^{(n)} & \Phi_{j,-i-1}^{(n)} \\
               \Phi_{-j-1,i}^{(n)} & \Phi_{-j-1,-i-1}^{(n)}
             \end{bmatrix},\;i,j\in\mathbb{Z}_{\geq0}.
\end{equation}
\begin{proof}
Since $\breve{\Phi}_{ji}=0_{2d^2}$ for $|i-j|>1,$ it is easy to see that \eqref{PHIbrevevshat} holds for $n=1.$ So, suppose that \eqref{PHIbrevevshat} is valid for some $n,$ then
\begin{eqnarray}
\nonumber  \breve{\Phi}_{ji}^{(n+1)} &=& [\breve{\Phi}\breve{\Phi}^{n}]_{ji}=\sum_{k=0}^{\infty}\breve{\Phi}_{jk}\breve{\Phi}_{ki}^{(n)}=\breve{\Phi}_{j,j-1}\breve{\Phi}_{j-1,i}^{(n)}+\breve{\Phi}_{jj}\breve{\Phi}_{ji}^{(n)}+\breve{\Phi}_{j,j+1}\breve{\Phi}_{j+1,i}^{(n)} \\
\nonumber   &=& M_{j-1}\breve{\Phi}_{j-1,i}^{(n)}+G_{j}\breve{\Phi}_{ji}^{(n)}+N_{j+1}\breve{\Phi}_{j+1,i}^{(n)}.
\end{eqnarray}
By the induction hypothesis and the result above,
\begin{eqnarray}
\nonumber &&\breve{\Phi}_{ji}^{(n+1)}=\\
\nonumber   && \begin{bmatrix}
         A_{j-1} & 0 \\
         0 & C_{-j}
       \end{bmatrix}
       \begin{bmatrix}
               \Phi_{j-1,i}^{(n)} & \Phi_{j-1,-i-1}^{(n)} \\
               \Phi_{-j,i}^{(n)} & \Phi_{-j,-i-1}^{(n)}
             \end{bmatrix}+
    \begin{bmatrix}
         B_{j} & 0 \\
         0 & B_{-j-1}
       \end{bmatrix}
       \begin{bmatrix}
               \Phi_{j,i}^{(n)} & \Phi_{j,-i-1}^{(n)} \\
               \Phi_{-j-1,i}^{(n)} & \Phi_{-j-1,-i-1}^{(n)}
             \end{bmatrix}\\
\nonumber   &&+ \begin{bmatrix}
         C_{j+1} & 0 \\
         0 & A_{-j-2}
       \end{bmatrix}
       \begin{bmatrix}
               \Phi_{j+1,i}^{(n)} & \Phi_{j+1,-i-1}^{(n)} \\
               \Phi_{-j-2,i}^{(n)} & \Phi_{-j-2,-i-1}^{(n)}
             \end{bmatrix} \\
\nonumber   &=&
       \begin{bmatrix}
               A_{j-1}\Phi_{j-1,i}^{(n)}+B_{j}\Phi_{j,i}^{(n)}+C_{j+1}\Phi_{j+1,i}^{(n)} & A_{j-1}\Phi_{j-1,-i-1}^{(n)}+B_{j}\Phi_{j,-i-1}^{(n)}+C_{j+1}\Phi_{j+1,-i-1}^{(n)} \\
               C_{-j}\Phi_{-j,i}^{(n)}+B_{-j-1}\Phi_{-j-1,i}^{(n)}+ A_{-j-2} \Phi_{-j-2,i}^{(n)} & C_{-j}\Phi_{-j,-i-1}^{(n)}+B_{-j-1}\Phi_{-j-1,-i-1}^{(n)}+A_{-j-2}\Phi_{-j-2,-i-1}^{(n)}
             \end{bmatrix}\\
\nonumber&=&\begin{bmatrix}
               \Phi_{ji}^{(n+1)} & \Phi_{j,-i-1}^{(n+1)} \\
               \Phi_{-j-1,i}^{(n+1)} & \Phi_{-j-1,-i-1}^{(n+1)}
             \end{bmatrix}.
\end{eqnarray}
\end{proof}
\end{pro}

Note that we can evaluate $\breve{\Phi}_{ji}^{(n)}$ by \eqref{KMFqmc} and then extract the block $\Phi_{ji}^{(n)}$ as in \eqref{PHIbrevevshat}. Further, for a density operator $\rho\in M_N(\mathbb{C}),$ we have
$$
p_{ji;\rho}(n)=\mathrm{Tr}\left(\Phi_{ji}^{(n)}\rho\right)=
\mathrm{Tr}\left(\begin{bmatrix}
\Phi_{ji}^{(n)} & 0 \\
 0 & 0
 \end{bmatrix}
\begin{bmatrix}
\rho  \\
0
\end{bmatrix}\right)
=\mathrm{Tr}\left(\begin{bmatrix}
I_{N^2} & 0 \\
0 & 0
\end{bmatrix}\breve{\Phi}_{ji}^{(n)}
\begin{bmatrix}
I_{N^2} & 0 \\
 0 & 0
 \end{bmatrix}
\begin{bmatrix}
\rho  \\
0
\end{bmatrix}\right).
$$
However, we would like to obtain the probability above avoiding the evaluation of $\breve{\Phi}_{ji}^{(n)}.$ This can be done via a generalization of the Karlin-McGregor formula on $\mathbb{Z}_{\geq0}$. We proceed as follows: first, write the decomposition
$$dW(x)=\begin{bmatrix}
                              dW_{11}(x) & dW_{12}(x) \\dW_{21}(x) & dW_{22}(x)
                              \end{bmatrix},$$
where $dW_{21}(x)=dW_{12}^*(x)$, since $dW(x)$ is positive definite. Then one has for $i,j\in\mathbb{Z}_{\geq0},$
\begin{eqnarray}
\nonumber\breve{\Phi}_{ji}^{(n)}&=&             \breve{\Pi}_j\int_\mathbb{R} x^n\mathcal{Q}_{j}^*(x)dW(x)\mathcal{Q}_{i}(x) \\
\nonumber   &\stackrel{\eqref{blobloQn}}{=}& \begin{bmatrix}\Pi_j & 0 \\0 & \Pi_{-j-1}\end{bmatrix}
\int_\mathbb{R} x^n\begin{bmatrix}
          Q_j^1(x) & Q_{-j-1}^1(x) \\
          Q_j^2(x) & Q_{-j-1}^2(x)
        \end{bmatrix}^*
\begin{bmatrix}
  dW_{11}(x) & dW_{12}(x) \\
  dW_{12}^*(x) & dW_{22}(x)
\end{bmatrix}
\begin{bmatrix}
          Q_i^1(x) & Q_{-i-1}^1(x) \\
          Q_i^2(x) & Q_{-i-1}^2(x)
        \end{bmatrix}
 \\
\nonumber   &=& \sum_{\alpha,\beta=1}^{2}
\begin{bmatrix}
\Pi_j\int_\mathbb{R}  x^nQ_j^{\alpha *}(x)dW_{\alpha\beta}(x)Q_i^{\beta}(x) & \Pi_j\int_\mathbb{R}  x^nQ_j^{\alpha *}(x)dW_{\alpha\beta}(x)Q_{-i-1}^{\beta }(x) \\
\Pi_{-j-1}\int_\mathbb{R}  x^nQ_{-j-1}^{\alpha *}(x)dW_{\alpha\beta}(x)Q_i^{\beta }(x) & \Pi_{-j-1}\int_\mathbb{R}  x^nQ_{-j-1}^{\alpha * }(x)dW_{\alpha\beta}(x)Q_{-i-1}^{\beta}(x)
\end{bmatrix} .
\end{eqnarray}
Joining equation above and Proposition \ref{Phibvsn}, we obtain the Karlin-McGregor formula for a QMC on $\mathbb{Z}$, given by
\begin{equation}\label{brevevsn}
\Phi_{ji}^{(n)}=\sum_{\alpha,\beta=1}^{2}\Pi_j\int_\mathbb{R} x^nQ_j^{\alpha *}(x)dW_{\alpha\beta}(x)Q_i^{\beta}(x),\;\;\mbox{for any}\; i,j\in \mathbb{Z},\;n\geq 0.
\end{equation}
Conversely, if there exist weight matrices $dW_{11}(x),dW_{12}(x),dW_{22}(x)$ such that $\Phi_{ji}^{(n)}$ is of the form \eqref{brevevsn}, then $\breve{\Phi}_{ji}^{(n)}$ is of the form
$$
\breve{\Phi}_{ji}^{(n)}=\breve{\Pi}_j\int_\mathbb{R} x^n\mathcal{Q}_{j}^*(x)dW(x)\mathcal{Q}_{i}(x).
$$
The weight matrix
\begin{equation*}\label{matrixspecphi}
W(x)=\begin{bmatrix}
            W_{11}(x) & W_{12}(x) \\
            W_{12}^*(x) & W_{22}(x)
          \end{bmatrix},
\end{equation*}
is called the spectral block matrix of $\Phi.$

\begin{remark}\label{remarkLIMstieltjes}
Extending Theorem \ref{OQWteostieltjes} to the QMC on $\mathbb{Z},$ we observe that, since $Q_0^1=Q_{-1}^2=I_N$ and $Q_0^2=Q_{-1}^1=0_N$, we obtain
\begin{eqnarray}
\nonumber   \sum_{n=0}^{\infty}p_{00;\rho}(n) &=& \sum_{n=0}^{\infty}\mathrm{Tr}\left[\Phi_{00}^{(n)}vec(\rho)\right]=\lim_{z\rightarrow 1}\sum_{n=0}^{\infty}z^n\mathrm{Tr}\left[\Pi_0\int_\mathbb{R}  x^nQ_0^{1*}(x)dW_{11}Q_0^{1}(x)vec(\rho)\right] \\
\nonumber   &=& \lim_{z\rightarrow 1}\sum_{n=0}^{\infty}\mathrm{Tr}\left[\Pi_0\int_\mathbb{R}  (zx)^n(x)dW_{11}(x)vec(\rho)\right]=\lim_{z\rightarrow 1}\mathrm{Tr}\left[\Pi_0\frac{dW_{11}(x)}{1-zx}vec(\rho)\right]\\
\nonumber&=&\lim_{z\rightarrow 1}z\;\mathrm{Tr}\left[\Pi_0B(z^{-1};W_{11})vec(\rho)\right]=\lim_{z\rightarrow 1}\;\mathrm{Tr}\left[\Pi_0B(z;W_{11})vec(\rho)\right],
\end{eqnarray}
where $B(z;W)$ is the Stieltjes transform of a weight matrix $W$ defined by \eqref{stieltjesDef}. Analogously,
$$
\sum_{n=0}^{\infty}p_{-1,-1;\rho}(n)=\lim_{z\rightarrow 1}\;\mathrm{Tr}\left[\Pi_{-1}B(z;W_{22})vec(\rho)\right].
$$
Since we are assuming that $\Pi_0$ and $\Pi_{-1}$ are positive definite matrices, vertex $\ket{0}$ is $\rho$-recurrent if and only if
\begin{equation*}\label{indBz0}
\lim_{z\downarrow 1}\mathrm{Tr}\left(B(z;W_{11})vec(\rho)\right)=\infty,
\end{equation*}
and
vertex $\ket{-1}$ is $\rho$-recurrent if and only if
\begin{equation*}\label{indBz1}
\lim_{z\downarrow 1}\mathrm{Tr}\left(B(z;W_{22})vec(\rho)\right)=\infty.
\end{equation*}
\end{remark}

Let us write the matrix $\Phi$ in the form
\begin{equation}\label{decmaismenos}
\Phi=\begin{bmatrix}
             \Phi^- & C \\
             A & \Phi^+
           \end{bmatrix},\;\;
C=\begin{bmatrix}
    \vdots & \vdots & \vdots &  \\
    0 & 0 & 0 & \cdots \\
    C_0 & 0 & 0 & \cdots
  \end{bmatrix},\;\;
A=\begin{bmatrix}
    \cdots & 0 & 0 & A_{-1} \\
    \cdots & 0 & 0 & 0 \\
    \cdots & 0 & 0 & 0 \\
     & \vdots & \vdots & \vdots
  \end{bmatrix},
\end{equation}
$$
\Phi^+=
\begin{bmatrix}
  B_0 & C_1 &  & & \\
  A_0&B_1 & C_2   &&  \\
  &A_1&B_2& C_3    & \\
   &  & \ddots &\ddots &\ddots
\end{bmatrix},\;\;
\Phi^-=
\begin{bmatrix}
  \ddots & \ddots & \ddots & & \\
  &A_{-4} & B_{-3}   &C_{-2}&  \\
  &&A_{-3}& B_{-2}&C_{-1} \\
   &  &  &A_{-2} &B_{-1}
\end{bmatrix}.
$$
Our goal now is to write the Stieltjes transforms associated with the weight matrices $W_{\alpha\beta},\alpha,\beta=1,2,$ in terms of the Stieltjes transforms associated with $W_{\pm}$, the weight matrices associated with $\Phi^{\pm}$. For that we will need the following lemma.
\begin{lemma}\cite{gv}\label{LemmaGV1}
Let $\mathcal{B}$ be a Banach space and $T_1:Dom(T_1)\rightarrow \mathcal{B}$ and $T_2:Dom(T_2)\rightarrow \mathcal{B}$ be linear operators with block representations
$$T_1=\begin{bmatrix}
  A & 0 \\
  C & D
\end{bmatrix}\;\;\mbox{ and }\;\;\;
T_2=\begin{bmatrix}
  A & C \\
  0 & D
\end{bmatrix},$$
respectively. If $A$ and $D$ are invertible, then $T_1$ and $T_2$ have inverses, given by
$$T_1^{-1}=\begin{bmatrix}
  A^{-1} & 0 \\
  -D^{-1}CA^{-1} & D^{-1}
\end{bmatrix}\;\;\mbox{ and }\;\;\;
T_2^{-1}=\begin{bmatrix}
  A^{-1} & -A^{-1}CD^{-1} \\
  0 & D^{-1}
\end{bmatrix}.$$
\end{lemma}
Denote by $\mathbb{P}_k,\;\mathbb{P}_k^-$ and $\mathbb{P}_k^+$ the projection maps onto the space generated by site $\ket{k}$ on $\mathbb{Z},\;\mathbb{Z}_{<0}$ and $\mathbb{Z}_{\geq0}$, respectively, and $\mathbb{Q}_k=I_{\mathbb{Z}}-\mathbb{P}_k,$ $\mathbb{Q}_k^-=I_{\mathbb{Z}_{<0}}-\mathbb{P}_k^-,$ $\mathbb{Q}_k^+=I_{\mathbb{Z}_{\geq0}}-\mathbb{P}_k^+$. Then, applying Lemma \ref{LemmaGV1}, we obtain
\begin{equation} \label{PhivsInv}
\begin{split}
\Phi(I-z\mathbb{Q}_0\Phi)^{-1} &= \begin{bmatrix}
             \Phi^- & C \\
             A & \Phi^+
           \end{bmatrix}
\begin{bmatrix}
  I-z\Phi^- & -zC \\
  0 & I-z\mathbb{Q}_0^+\Phi^+
\end{bmatrix}^{-1} \\
&=\begin{bmatrix}
             \Phi^- & C \\
             A & \Phi^+
           \end{bmatrix}
\begin{bmatrix}
                                   (I-z\Phi^-)^{-1} & z(I-z\Phi^-)^{-1}C(I-z\mathbb{Q}_0^+\Phi^+)^{-1} \\
                                   0 & (I-z\mathbb{Q}_0^+\Phi^+)^{-1}
                                 \end{bmatrix}
\\
&=\begin{bmatrix}
  \Phi^-(I-z\Phi^-)^{-1} & \left[z\Phi^-(I-z\Phi^-)^{-1}+I\right]C(I-z\mathbb{Q}_0^+\Phi^+)^{-1} \\
  A(I-z\Phi^-)^{-1} & [zA(I-z\Phi^-)^{-1}C+\Phi^+](I-z\mathbb{Q}_0^+\Phi^+)^{-1}
\end{bmatrix}.
\end{split}
\end{equation}
By the same arguments,
\begin{equation*} \label{PhivsInv2}
\begin{split}
\Phi(I-z\mathbb{Q}_{-1}\Phi)^{-1} &= \begin{bmatrix}
             \Phi^- & C \\
             A & \Phi^+
           \end{bmatrix}
\begin{bmatrix}
  I-z\mathbb{Q}_{-1}^-\Phi^- & 0 \\
  -zA &I-z\Phi^+
\end{bmatrix}^{-1} \\
&=\begin{bmatrix}
             \Phi^- & C \\
             A & \Phi^+
           \end{bmatrix}
\begin{bmatrix}
  (I-z\mathbb{Q}_{-1}^-\Phi^-)^{-1} & 0 \\
  z(I-z\Phi^+)^{-1}A(I-z\mathbb{Q}_{-1}^-\Phi^-)^{-1} &(I-z\Phi^+)^{-1}
\end{bmatrix}
\\
&=\begin{bmatrix}
  (\Phi^-+zC(I-z\Phi^+)^{-1}A)(I-z\mathbb{Q}_{-1}^-\Phi^-)^{-1} & C(I-z\Phi^+)^{-1} \\
  \left(I+z\Phi^+(I-z\Phi^+)^{-1}\right)A(I-z\mathbb{Q}_{-1}^-\Phi^-)^{-1} & \Phi^+(I-z\Phi^+)^{-1}
\end{bmatrix},
\end{split}
\end{equation*}
and
$$
C(I-z\mathbb{Q}_0\Phi^+)^{-1}=\begin{bmatrix}
    \vdots & \vdots & \vdots &  \\
    0 & 0 & 0 & \cdots \\
    C_0 & 0 & 0 & \cdots
  \end{bmatrix}
\begin{bmatrix}
  I & 0 \\
  * & *
\end{bmatrix}^{-1}
=\begin{bmatrix}
   \vdots & \vdots &  \\
   0 & 0 & \cdots \\
   C_0 & 0 & \cdots
 \end{bmatrix}.
$$
Denoting
$$\Phi^-(z):=\sum_{n=0}^{\infty}z^n\left(\Phi^-\right)^n=(I-z\Phi^-)^{-1},\quad \Phi^+(z):=\sum_{n=0}^{\infty}z^n\left(\Phi^+\right)^n=(I-z\Phi^+)^{-1},$$
we obtain
\begin{eqnarray}
\nonumber  F_{00}(z) &=& z\mathbb{P}_0\Phi(I-z\mathbb{Q}_0\Phi)^{-1}\mathbb{P}_0 \\
\nonumber   &=& \begin{bmatrix}
                  0 & 0 \\
                  0 & z\mathbb{P}_0^+\left[zA(I-z\Phi^-)^{-1}C(I-z\mathbb{Q}_0\Phi^+)^{-1}+\Phi^+(I-z\mathbb{Q}_0^+\Phi^+)^{-1}\right]\mathbb{P}_0^+
                \end{bmatrix},
\end{eqnarray}
where the only non-null block equals
\begin{align*}
=&z^2\mathbb{P}_0^+\left[
\begin{bmatrix}
    \cdots &  A_{-1}\Phi^{-}_{-1,-2}(z) & A_{-1}\Phi^{-}_{-1,-1}(z) \\
    \cdots &  0 & 0 \\
    \cdots &  0 & 0 \\
     &  \vdots & \vdots\end{bmatrix}
\begin{bmatrix}
    \vdots & \vdots & \vdots &  \\
    0 & 0 & 0 & \cdots \\
    C_0 & 0 & 0 & \cdots
  \end{bmatrix}
\right]\mathbb{P}_0^+ +F_{00}^+(z)\\
&= z^2\mathbb{P}_0^+
\begin{bmatrix}
  A_{-1}\Phi^{-}_{-1,-1}(z)C_0 & 0 \\
  0 & 0
\end{bmatrix}\mathbb{P}_0^+ +F_{00}^+(z)= z^2
\begin{bmatrix}
  A_{-1}\Phi^{-}_{-1,-1}(z)C_0 & 0 \\
  0 & 0
\end{bmatrix}+F_{00}^+(z).
\end{align*}
Note that $F_{00}(z)$ has only one non-null $N^2\times N^2$ block, due to the projections multiplying on the left and on the right-hand side. Without loss of generality, we will rewrite this kind of blocks as its only non-null block. For instance, we have
\begin{equation*}\label{F00}
F_{00}(z)=z^2A_{-1}\Phi^{-}_{-1,-1}(z)C_0+F_{00}^+(z).
\end{equation*}
Applying twice the equation
\beq\label{oldeq100}  F_{ji}(s) = \Phi_{jj}(s)^{-1}(\Phi_{ji}(s)-\delta_{ji}I),\eeq
for $F_{00}(z)$ and $F^+_{00}(z),$ we obtain
$$
 I-\Phi_{00}(z)^{-1}=z^2A_{-1}\Phi^{-}_{-1,-1}(z)C_0 +I-\Phi_{00}^+(z)^{-1},
$$
and after some algebra, we get
\begin{equation}\label{PP1}
\Phi_{00}(z)=\Phi_{00}^+(z)(I-z^2A_{-1}\Phi_{-1,-1}^-(z)C_0\Phi_{00}^+(z))^{-1}.
\end{equation}
Analogously,
\begin{align*}
 F_{-1,-1}(z) &= z\mathbb{P}_{-1}^-\left[\Phi^-(I-z\mathbb{Q}_{-1}\Phi^-)^{-1}+zC(I-z\Phi^+)^{-1}A(I-z\mathbb{Q}_0\Phi^-)^{-1}\right]
\mathbb{P}_{-1}^- \\
&= F_{-1,-1}^-(z)+z^2C\Phi_{00}^+(z)A_{-1},
\end{align*}
thus
\begin{align*}
\Phi_{-1,-1}(z) &= (I-F_{-1,-1}(z))^{-1} =(I-F_{-1,-1}^-(z)-z^2C\Phi_{00}^+(z)A_{-1})^{-1}\\
&= \Phi^-_{-1,-1}(z)(I-z^2C_0\Phi^+_{00}(z)A_{-1}\Phi_{-1,-1}^-(z))^{-1},
\end{align*}
that is,
\begin{equation}\label{PP2}
\Phi_{-1,-1}(z)=\Phi^-_{-1,-1}(z)(I-z^2C_0\Phi^+_{00}(z)A_{-1}\Phi_{-1,-1}^-(z))^{-1}.
\end{equation}
Now we use equation \eqref{PhivsInv} to obtain
$$
F_{0,-1}(z)=z\mathbb{P}_0A(I-z\Phi^-)^{-1}\mathbb{P}_{-1}=zA_{-1}\Phi_{-1,-1}^-(z),
$$
which, together with equations \eqref{oldeq100} and \eqref{PP1}, gives
\begin{equation}\label{PP3}
\Phi_{0,-1}(z) = \Phi_{00}(z)F_{0,-1}(z)= z\Phi_{00}^+(z)(I-z^2A_{-1}\Phi_{-1,-1}^-(z)C_0\Phi_{00}^+(z))^{-1}A_{-1}\Phi_{-1,-1}^-(z).
\end{equation}
In the same way,
$$
F_{-1,0}(z)=zC_0\Phi_{00}^+(z),
$$
gives
\begin{equation}\label{PP4}
\Phi_{-1,0}(z) = \Phi_{-1,-1}(z)F_{-1,0}(z)= z\Phi^-_{-1,-1}(z)(I-z^2C_0\Phi^+_{00}(z)A_{-1}\Phi_{-1,-1}(z))^{-1}C_0\Phi_{00}^+(z).
\end{equation}

We notice that the block matrices of both $\Phi^+$ and $\Phi^-$ satisfy the conditions of equation \eqref{condiequiv}, thus there are positive weight matrices $W_\pm$ associated with $\Phi^\pm$ for which the associated polynomials are orthogonal. Then, we can write
$$
\Pi_0^+:=\int_\mathbb{R}dW_+\;\;\;\mbox{ and }\;\;\;\Pi_{-1}^-:=\int_\mathbb{R}dW_-\;\;.
$$
Recalling that (see \eqref{genfunct})
$$\Phi_{ji}(s) = \Pi_j\int_\mathbb{R} \frac{1}{1-sx}Q_j^*(x)dW(x)Q_i(x),$$
and $Q_0^1=Q_{-1}^2=I_{N^2},$ $Q_0^2=Q_{-1}^1=0_{N^2}$, we obtain the following Stieltjes transforms relations
$$
\begin{array}{lll}
  B(z^{-1};W_{11})=z\Pi_0^{-1}\Phi_{00}(z),&B(z^{-1};W_{22})=z\Pi_{-1}^{-1}\Phi_{-1,-1}(z),&B(z^{-1};W_{12})=z\Pi_{-1}^{-1}\Phi_{0,-1}(z), \\
  B(z^{-1};W_{21})=z\Pi_{-1}^{-1}\Phi_{-1,0}(z),&B(z^{-1};W_+)=z(\Pi^{+}_0)^{-1}\Phi_{00}^+(z),&B(z^{-1};W_-)=z(\Pi^{-}_{-1})^{-1}\Phi_{-1,-1}^-(z).
\end{array}
$$
Joining with the identities \eqref{PP1},\eqref{PP2},\eqref{PP3},\eqref{PP4},  the new Stieltjes transform identities are obtained:
\begin{equation}\label{newstieltjes}
\begin{split}
\Pi_{0} B(z;W_{11}) &= \Pi_{0}^+B(z;W_+)(I-A_{-1}\Pi_{-1}^-B(z;W_-)C_0\Pi_{0}^+B(z;W_+))^{-1}, \\
\Pi_{-1} B(z;W_{22})&=\Pi_{-1}^-B(z;W_-)(I-C_0\Pi_{0}^+B(z;W_+)A_{-1}\Pi_{-1}^-B(z;W_-))^{-1},  \\
 \Pi_{0} B(z;W_{12}) &= \Pi_{0}^+B(z;W_+)(I-A_{-1}\Pi_{-1}^-B(z;W_-)C_0\Pi_{0}^+B(z;W_+))^{-1}A_{-1}\Pi_{-1}^-B(z;W_-), \\
 \Pi_{-1} B(z;W_{21}) &=\Pi_{-1}^-B(z;W_-)(I-C_0\Pi_{0}^+B(z;W_+)A_{-1}\Pi_{-1}^-B(z;W_-))^{-1}C_0\Pi_{0}^+B(z;W_+).
\end{split}
\end{equation}
Sometimes the operators $\Pi_i^+$ and $\Pi_i^-$ are equal to the identity operator. In this case, \eqref{newstieltjes} are reduced to
\beq\label{newstieltjesORTH}
\begin{split}
\Pi_{0} B(z;W_{11}) &= B(z;W_+)(I-A_{-1}B(z;W_-)C_0B(z;W_+))^{-1}, \\
\Pi_{-1} B(z;W_{22})&=B(z;W_-)(I-C_0B(z;W_+)A_{-1}B(z;W_-))^{-1},  \\
\Pi_{0} B(z;W_{12}) &=  B(z;W_+)(I-A_{-1}B(z;W_-)C_0B(z;W_+))^{-1}A_{-1}B(z;W_-), \\
\Pi_{-1} B(z;W_{21}) &= B(z;W_-)(I-C_0B(z;W_+)A_{-1}B(z;W_-))^{-1}C_0B(z;W_+).
\end{split}
\eeq
The above results will be applied in the following examples so that one is able to conclude recurrence properties of the walk.
\bex
Let $\Phi$ be a homogeneous OQW on $\mathcal{S}=\mathbb{Z}$ with matrix representation
$$\Phi=
\left[\begin{array}{ccc|ccccc}
\ddots & \ddots &  &  &  &  &  &  \\
  \ddots & 0 & \lceil L \rceil  &  &  &  &  &  \\
   &\lceil R \rceil  & 0 & \lceil L \rceil &  &  &  &  \\
  \hline
   &    & \lceil R \rceil  & 0 & \lceil L \rceil & & &  \\
    &  &    & \lceil R \rceil  & 0 &\lceil L \rceil & & \\
    & & &    & \lceil R \rceil  & 0 &\lceil L \rceil &  \\
     &  &  &  &  & \ddots & \ddots &\ddots
\end{array}\right],\quad
R=\begin{bmatrix}
    \frac{1}{\sqrt{3}} & 0 \\
    0 & \frac{1}{\sqrt{2}}
  \end{bmatrix},\;\;
L=\begin{bmatrix}
    \frac{\sqrt{2}}{\sqrt{3}} & 0 \\
    0 & \frac{1}{\sqrt{2}}
  \end{bmatrix}.
$$
In order to study recurrence or transience of the walk for each density operator on $\mathbb{C}^2,$ we will apply the Stieltjes transformation discussed above. The polynomials associated with $\Phi$ are
\begin{eqnarray}
\nonumber  Q_0^1(x) &=& I_4, \;\; Q_0^2(x)=0_4\\
\nonumber  Q_{-1}^1(x) &=& 0_4,\;\; Q_{-1}^2(x)=I_4 \\
\nonumber   xQ_n^\alpha (x) &=& Q_{n+1}^\alpha(x)\lceil R \rceil +Q_{n-1}^\alpha(x)\lceil L \rceil ,\;\;\alpha,\beta=1,2,\;\;n\in\mathbb{Z}.
\end{eqnarray}

The weight matrix associated with $\Phi^+$ is
$$
W_+(x)=\begin{bmatrix}
  \frac{3\sqrt{2}}{4\pi}\left[\sqrt{\left(4-\frac{9x^2}{2}\right)}\right]_+  &&&  \\
        &\frac{2^{1/4}\sqrt{3}}{2\pi}\left[\sqrt{\left(\sqrt{2}(2\sqrt{2}-3x^2)\right)}\right]_+&&  \\
            &&\frac{2^{1/4}\sqrt{3}}{2\pi}\left[\sqrt{\left(\sqrt{2}(2\sqrt{2}-3x^2)\right)}\right]_+&  \\
                &&&\frac{2(x^2-1+\sqrt{1-x^2})}{x^2(1-x^2)}
  \end{bmatrix}
$$
and since the matrices are diagonal, it is easy to see that $W_+(x)=W_-(x).$ The weight matrix $W_{11}(x)$ is obtained by an application of the first formula of \eqref{newstieltjes},
$$
B(z;W_{11}) = B(z;W_+)(I-A_{-1}B(z;W_+)C_0B(z;W_+))^{-1},
$$
and then we apply the Perron-Stieltjes inversion formula to obtain the referred measure. After some calculus, we have, for a density matrix $\rho=\begin{bmatrix}
                                          a & b \\
                                          b^* & 1-a
                                        \end{bmatrix}$
on $\mathbb{C}^2,$
\begin{align*}
\sum_{n=0}^{\infty}p_{00;\rho}(n)  &=  \sum_{n=0}^{\infty}\mathrm{Tr}\left({\Phi}^{(n)}_{00}vec(\rho)\right)=\lim_{z\rightarrow\infty}\mathrm{Tr}\left({\Phi}_{00}(z)vec(\rho)\right)=\lim_{z\rightarrow\infty}\mathrm{Tr}\left(B(W_{11},z)vec(\rho)\right) \\
&\stackrel{\eqref{PP1}}{=}\lim_{z\rightarrow\infty}\frac{1-a}{\sqrt{1-z^2}}+\frac{6a(8\sqrt{2}z^2+3\sqrt{18-16z^2}-9\sqrt{2})}{(3\sqrt{2}+\sqrt{18-16z^2})(18-16z^2)}= \begin{cases}
                  \infty, & \mbox{ if }\; a<1 \\
                  3, & \mbox{ if }\; a=1
                \end{cases}.
\end{align*}
Therefore site $\ket{0}$ is $\rho$-transient for $\rho=\begin{bmatrix}
                                          1 & 0\\
                                          0 & 0
                                        \end{bmatrix}$
and $\rho$-recurrent otherwise.
\eex
\qee

It is worth recalling that the weight matrix of the example above is a particular case of Proposition 1.3 of \cite{jl}.

\bex\label{preqmcreta}
Consider a QMC $\hat{\Phi}$ induced by the block matrix on $V=\left\{0,1,2,\ldots\right\}$ given by
\begin{equation*}
\Phi=
\begin{bmatrix}
 B & rI &&&  \\
 tI & B & rI&& \\
 & tI & B & rI& \\
&  &  \ddots &\ddots&\ddots
\end{bmatrix}
,\;\;\;0<r,t<1,
\end{equation*}
where $B=[\sigma_B]$, $\sigma_B=V_1^*\cdot V_1+V_2^*\cdot V_2$, where $V_1$ and $V_2$ are the same as in the example appearing in Section \ref{sec5}. For simplicity we assume $0<a,b,s<1$, $a^2+b^2<1$. In this way we have that $\mathrm{Tr}(\sigma(X))=s\mathrm{Tr}(X)$, so we suppose that $r+s+t=1$ in order to have that $\hat{\Phi}$ is trace-preserving. The matrices $R_n=\left(\sqrt{\frac{r}{t}}\right)^n$ satisfy the conditions of Equation \eqref{condiequiv}, thus we denote
$$
\Pi_n=R_n^*R_n=\left(\frac{r}{t}\right)^n.
$$
By the classical symmetrization
$$\mathcal{Y}=diag(Y_0,Y_1,\ldots),\;\;\;Y_i=\Bigg(\sqrt{\frac{r}{t}}\Bigg)^{i-1}I_4,\;\;\;i=0,1,\ldots,$$
we obtain
$$
J=\mathcal{Y}\Phi\mathcal{Y}^{-1}=\begin{bmatrix} B & kI &  & & \\ kI & B & kI  & & \\  & kI & B & kI  & \\ && \ddots & \ddots & \ddots
\end{bmatrix},\;\;\;k=\sqrt{rt}.
$$
The matrix $B$ is symmetric, thus we can apply the spectral theorem to get
$$
B=UDU^*,\quad  D=s\begin{bmatrix}
                1 & 0 & 0 & 0 \\
                0 & 1 & 0 & 0 \\
                 0& 0 & 1-2a^2-2b^2 & 0 \\
                 0& 0 & 0 & 1-2a^2-2b^2
              \end{bmatrix},
$$
where
$$
U=\frac{\sqrt{2}}{2}\begin{bmatrix}
    1 & \frac{a}{\sqrt{a^2+b^2}} & -\frac{b}{\sqrt{2a^2+b^2}} & -\frac{ab}{\sqrt{2a^2+b^2}\sqrt{a^2+b^2}} \\
    0 & \frac{b}{\sqrt{a^2+b^2}} & \frac{2a}{\sqrt{2a^2+b^2}} & -\frac{b^2}{\sqrt{2a^2+b^2}\sqrt{a^2+b^2}} \\
    0 & \frac{b}{\sqrt{a^2+b^2}} & 0 & \frac{\sqrt{2a^2+b^2}}{\sqrt{a^2+b^2}} \\
    1 & -\frac{a}{\sqrt{a^2+b^2}} & \frac{b}{\sqrt{2a^2+b^2}} & \frac{ab}{\sqrt{2a^2+b^2}\sqrt{a^2+b^2}}
  \end{bmatrix},
$$
which gives
$$
H(x):=U
\begin{bmatrix}
  \frac{(s-x)^2}{k^2}-4 & 0 & 0 & 0 \\
 0 & \frac{(s-x)^2}{k^2}-4 & 0 & 0 \\
 0& 0 & \frac{(s(1-2a^2-2b^2)-x)^2}{k^2}-4 & 0 \\
 0& 0 & 0 & \frac{(s(1-2a^2-2b^2)-x)^2}{k^2}-4
\end{bmatrix}
U^*,
$$
and then the associated weight matrix is (\cite{duran})
$$
\begin{array}{l}
  dW(x)=\dfrac{1}{4\pi k(a^2+b^2)}\times \\
  \left(\left[w_1(x)\right]_+ \begin{bmatrix}
    2a^2+b^2 & ab & ab & b^2 \\
    ab & b^2 & b^2 & -ab \\
    ab & b^2 & b^2 & -ab \\
    b^2 & -ab & -ab & 2a^2+b^2
  \end{bmatrix}+
 \left[w_2(x)\right]_+ \begin{bmatrix}
  b^2 & -ab & -ab & -b^2 \\
  -ab & 2a^2+b^2 & -b^2 & ab \\
 -ab & -b^2 & 2a^2+b^2 & ab \\
  -b^2 & ab & ab & b^2
\end{bmatrix}\right)dx,
\end{array}
$$
where
$$
w_1(x)=\sqrt{4-\frac{(s-x)^2}{k^2}},\quad w_2(x)=\sqrt{4-\frac{(s(1-2a^2-2b^2)-x)^2}{k^2}}.
$$
Note that we can rewrite the weight matrix in terms of $w_1(x),w_2(x)$ and $B$ by
\begin{equation}\label{dWforB}
\begin{array}{rl}
  dW(x)= & \dfrac{w_1(x)}{4\pi k(a^2+b^2)}\left((2a^2+2b^2-1)I_4+\frac{1}{s}B\right)+\dfrac{w_2(x)}{4\pi
  k(a^2+b^2)}\left(I_4-\frac{1}{s}B\right) \\
   =&\dfrac{1}{2k\pi} U\begin{bmatrix}
        \left[w_1(x)\right]_+ &  &  &  \\
        & \left[w_1(x)\right]_+ &  &  \\
        &  & \left[w_2(x)\right]_+ &  \\
        &  &  &\left[w_2(x)\right]_+
      \end{bmatrix}U^*,
\end{array}
\end{equation}
whose support is given by
\begin{equation}\label{supprrr}
\begin{split}
  R:=supp(dW) &=\{y\in\mathbb{R}:\frac{1}{k}(yI_4-B)\mbox{ has an eigenvalue in }[-2,2]\} \\
   &=[-2k+s(1-2a^2-2b^2),s+2k].
\end{split}
\end{equation}
The Stieltjes transform of $W$ is
\begin{equation}\label{bzwex}
B(z;W)=\int_R\dfrac{1}{2k\pi} U\begin{bmatrix}
        \frac{w_1(x)}{z-x} &  &  &  \\
        & \frac{w_1(x)}{z-x} &  &  \\
        &  & \frac{w_2(x)}{z-x} &  \\
        &  &  &\frac{w_2(x)}{z-x}
      \end{bmatrix}U^*dx,
\end{equation}
where the integrals of the elements on the diagonal are
\begin{equation}\label{hhh}
\begin{split}
\int_R\frac{w_1(x)dx}{z-x}&=\frac{\pi}{k}(z-s-i\sqrt{4k^2-(s-z)^2}):=2k\pi h_1(z),\\
\int_R\frac{w_2(x)dx}{z-x}&=\frac{\pi}{k}(z-s(1-2a^2-2b^2)-i\sqrt{4k^2-(s(1-2a^2-2b^2)-z)^2}):=2k\pi h_2(z).
\end{split}
\end{equation}
The transience of this walk can be computed by using Theorem \ref{stieltjesteorec}:
{\color{black}\begin{eqnarray}
 \nonumber \lim_{z\downarrow 1}Tr\left[z\; vec^{-1}\left(B(z;W)vec\left(\begin{bmatrix}
                                                               u & v \\
                                                               v^* & 1-u
                                                             \end{bmatrix}\right)\right)\right] &=& \frac{1-s+\sqrt{s^2-2s+1-4k}}{2k^2} \\
\nonumber   &=& \frac{r+t+\sqrt{r^2-2rt+t^2}}{2rt} \\
\nonumber   &=& \begin{cases}
                  1/r, & \mbox{if } t\geq r \\
                  1/t, & \mbox{otherwise}.
                \end{cases}
\end{eqnarray}}

Since this limit is valid for any density operator $\rho=\begin{bmatrix}u & v \\v^* & 1-u\end{bmatrix}\in \mathbb{M}(\mathbb{C}^2),$ we conclude that
this QMC is transient.

\medskip

Let us extend the above QMC to the real line: now the set of vertices is $V=\mathbb{Z}$ and the new QMC $\Phi$ has matrix representation
$$
\Phi=
\begin{bmatrix}
  \ddots & \ddots & \ddots &  &&&  \\
   & tI & B & rI &&&  \\
   &  & tI & B & rI&& \\
      &  && tI & B & rI& \\
   &  &  &  &\ddots &\ddots&\ddots
\end{bmatrix}.
$$

Take the splitting of equation \eqref{decmaismenos} applied to $\Phi:$
$$\Phi=\begin{bmatrix}
             \Phi^- & C \\
             A & \Phi^+
           \end{bmatrix},\;\;
C=\begin{bmatrix}
    \vdots & \vdots & \vdots &  \\
    0 & 0 & 0 & \cdots \\
    rI & 0 & 0 & \cdots
  \end{bmatrix},\;\;
A=\begin{bmatrix}
    \cdots & 0 & 0 & tI \\
    \cdots & 0 & 0 & 0 \\
    \cdots & 0 & 0 & 0 \\
     & \vdots & \vdots & \vdots
  \end{bmatrix}.
$$
The weight matrix associated with $\Phi^+$ is $W_+=W$, where $W$ is given by \eqref{dWforB} and with support $R$ given by \eqref{supprrr}. We have $\Pi_0^+=\Pi_{-1}^-=I_4$ and the Stieltjes transform of $W_+$ is given by \eqref{bzwex} and \eqref{hhh}. The operators $\Pi_0=R_0^*R_0$ and $\Pi_{-1}=R_{-1}^*R_{-1}$ are the ones obtained by equation \eqref{condiequiv}, giving $\Pi_0=I$ and $\Pi_{-1}=A^{-1}C=\frac{r}{t}I.$ For simplicity, assume $s=2k$. Then, we apply formula \eqref{newstieltjes} to obtain
$$
B(z;W_{11})=U\begin{bmatrix}
        l_1(z) &  &  &  \\
        & l_1(z) &  &  \\
        &  & l_2(z) &  \\
        &  &  &l_2(z)
      \end{bmatrix}U^*,
$$
where
$$
l_1(z)=\frac{\sqrt{z(4k-z)}}{z(z-4)},\;l_2(z)=\frac{\sqrt{-z(z+4k)}}{z(4k-z)},
$$
and we evaluate
\begin{align*}
B(z;W_{22}) &= \frac{t}{r}B(z;W_{11}) \\
B(z;W_{21})= B(z;W_{12}) &= tB(z;W_{11})B(z;W_+)=
tU\begin{bmatrix}
        h_1(z)l_1(z) &  &  &  \\
        & h_1(z)l_1(z) &  &  \\
        &  & h_2(z)l_2(z) &  \\
        &  &  &h_2(z)l_2(z)
      \end{bmatrix}U^*,
\end{align*}
where $h_i(z),i=1,2$ are defined by \eqref{hhh}. Applying  [\cite{dIbook}, eq. (1.10)] we obtain the spectral measure of $\Phi$,
$$
dW(x)=\begin{bmatrix}
             U & 0 \\
             0 & U
           \end{bmatrix}
\begin{bmatrix}
  D_{11}(x) &  D_{12}(x)\\
  D_{12}(x) & \frac{t}{r}D_{11}(x)
\end{bmatrix}
\begin{bmatrix}
 U^* & 0 \\
  0 & U^*
\end{bmatrix},
$$
where
$$
 D_{11}(x)=diag\left(\dfrac{-1}{\left[\sqrt{x(4k-x)}\right]_+},\dfrac{-1}{\left[\sqrt{x(4k-x)}\right]_+},\dfrac{-1}{\left[\sqrt{-x(4k+x)}\right]_+},\dfrac{-1}{\left[\sqrt{-x(4k+x)}\right]_+}\right),
$$
$$
D_{12}(x)=diag\left(\dfrac{2k-x}{2r\left[\sqrt{x(4k-x)}\right]_+},\dfrac{2k-x}{2r\left[\sqrt{x(4k-x)}\right]_+},
\dfrac{-2k-x}{2r\left[\sqrt{-x(4k+x)}\right]_+},\dfrac{-2k-x}{2r\left[\sqrt{-x(4k+x)}\right]_+}\right).
$$
The procedure to obtain the spectral measure for $\Phi$ was inspired by the classical case. The reader can note that the expressions appearing in (\ref{newstieltjesORTH}) are analogous to the classical reasoning. However, some of the transition matrices do not commute, thus the order of the operators in such formulae has to be maintained.

\medskip

Now, for any density operator on $\mathbb{C}^2,$ we have by Remark \ref{remarkLIMstieltjes} that
$$
\sum_{n=0}^{\infty}p_{00;\rho}(n) = \lim_{z\rightarrow 1}\mathrm{Tr}\left(\Pi_0^{-1}B(z;W_{11})vec(\rho)\right)= \lim_{z\rightarrow 1}\frac{1}{\sqrt{z(z-4k)}}=\begin{cases}
                             \frac{1}{\sqrt{1-4k}}, & \mbox{if } \quad k<1/4, \\
                             \infty, & { if }\quad k=1/4.
                           \end{cases}
$$
That is, the walk $\Phi$ (for $s=2k$) is recurrent only when $k=1/4$ and this happens for $t=r=1/4.$ For the general case we can follow the same steps to obtain
$$
\sum_{n=0}^{\infty}p_{00;\rho}(n)=\lim_{z\rightarrow 1}\frac{1}{\sqrt{z^2-2sz+s^2-4k^2}}=\begin{cases}
                             \frac{1}{\sqrt{1-2s+s^2-4k^2}}, & \mbox{if } s\neq 1-2k, \\
                             \infty, & { if }s=1-2k.
                           \end{cases}
$$
Since we are assuming $r+s+t=1$ and $k=\sqrt{rt},$ recurrence occurs when $0=r-2\sqrt{rt}+t=(\sqrt{r}-\sqrt{t})^2,$ that is, when $t=r.$
\eex
\qee

\begin{remark}
The example in Section \ref{sec5} is such that $\sigma_B+t^2I< I,$ thus $\sum_{j=0}^{\infty}p_{0j;\rho}(n)<1$ for some initial density operator $\rho.$ This case is interpreted as a walk with a vertex named $\ket{-1},$ which is an absorbing vertex of the QMC, giving the correction $\sum_{j=-1}^{\infty}p_{0j;\rho}(n)=1.$
Now we point out the difference that an absorbing vertex on the QMC can take: the QMC $\Phi$ acting on $\mathbb{Z}_{\geq 0}$ has an absorbing vertex on site $\ket{0},$ and it is transient for any choice of $t,r,s,a,b$. On the other hand, for $a,b,s$ fixed and  $t=r=1-s$, the extended QMC on the integer line is always recurrent.
\end{remark}

\section{The case of non-symmetric weight matrices}\label{sec8}

As discussed previously, Theorem \ref{dette 2.1} describes the fundamental conditions regarding the existence of a positive weight matrix associated with a given QMC. Then, a natural question arises: is there anything that can be done in the case of QMC that do not satisfy such conditions, perhaps involving a non symmetric matrix of measures? Based on \cite{zyge2}, we are in fact able to discuss a non-general Karlin-McGregor formula for $\Phi$ by using a different
kind of polynomial orthogonality, where the term $\emph{non-general}$ means that we obtain the $(i,j)$-th block entry of $\Phi^n$ only for $i=0,$ which will allow us to obtain certain developments for the recurrence problems we are interested in.

\medskip

We will be mostly interested in homogeneous QMCs, that is, operators $\Phi$ of the form $\eqref{phirenewaleq},$ such that $A_n=A, B_n=B, C_{n+1}=C,\;\forall
n=0,1,2,\ldots$ for some $A,B,C\in M_{N^2}(\mathbb{C}).$ For instance, if we have a homogeneous OQW with
\begin{equation*}\label{ABCattal}
A=\frac{1}{\sqrt{3}}\begin{bmatrix}
     1 & 0 \\
     -1 & 1
   \end{bmatrix},\quad
C=\frac{1}{\sqrt{3}}\begin{bmatrix}
     1 & 1 \\
     0 & 1
   \end{bmatrix},\quad
B=0_2,
\end{equation*}
then $A_0C_1$ is not Hermitian, consequently it is not possible to obtain a proper positive definite weight matrix $W$ that makes the corresponding matrix-valued polynomials orthogonal with respect to $W.$ However, we may consider another kind of orthogonality for the associated polynomials in terms of a reasoning seen in \cite{zyge2}. For a homogeneous
QMC, Theorem 3.4 of \cite{zyge2} assures the existence of a weight matrix $W$ supported on some subspace $\Delta$ of $\mathbb{C}$ such that
the polynomials $Q_n(x),$ defined recursively by
\begin{equation}\label{3thermmatrixb}
\begin{split}
Q_0(x)&=I_{N^2},\quad Q_{-1}(x)=0_{N^2}, \\
  xQ_n(x) &= Q_{n+1}(x)A_{n}+Q_n(x)B_n+Q_{n-1}(x)C_{n},
\end{split}
\end{equation}
satisfy
\begin{equation}\label{zygortohogoanlity}
\int_\Delta x^kdW(x)Q_n(x)=0,
\end{equation}
for all integers $n>k\geq 0.$ Polynomials $\{Q_n(x)\}_{n\geq0}$ for which there exists a weight matrix $W$ satisfying \eqref{zygortohogoanlity} are
called \textbf{semi-orthogonal polynomials}\index{semi-orthogonal polynomials} with respect to $W$. Since this concept of orthogonality is
weaker, the Karlin-McGregor formula for non-symmetric QMCs will be weaker as well. Nevertheless, we will be able to obtain an application of such construction for the problem of recurrence.

\medskip

For completeness, let us derive the Karlin-McGregor formula for non-symmetric weight matrices with the necessary adaptations with respect to semi-orthogonality. We have $x^nQ(x)=Q(x)\Phi^n,$ where $Q(x)=(Q_0(x),Q_1(x),\ldots).$ Component-wise,
\begin{equation}\label{cwiseb}
x^nQ_r(x)=\sum_{k=0}^{\infty}Q_k(x)\Phi_{kr}^{(n)}.
\end{equation}
Fix $i,j\in \mathbb{Z}_{\geq0}$ vertices. Fix a time parameter $n$ with the extra condition $n\geq i,$ then multiply $Q_j^*(x)$ on the left-hand side of
\eqref{cwiseb} with $r=j+i$ and integrate on $\Delta$ to obtain
\begin{equation}\label{preKMGzyg}
\int_\Delta x^nQ_j^*(x)dW(x)Q_{j+i}(x)=\sum_{k=0}^{\infty}\int_\Delta Q_j^*(x)dW(x)Q_k(x)\Phi_{k,j+i}^{(n)}\stackrel{\eqref{zygortohogoanlity}}{=}\sum_{k=0}^{j}\int_\Delta Q_j^*(x)dW(x)Q_k(x)\Phi_{k,j+i}^{(n)}.
\end{equation}
Hypothesis $n<i$ in this situation would make the integral on the left-hand side of \eqref{preKMGzyg} to vanish, by an application of
$\eqref{zygortohogoanlity}.$ The same idea is applied to the right-hand side of \eqref{preKMGzyg}, where we want the sum of integrals to become only
one term, which happens for the particular case $j=0$:
$$
\int_\Delta x^nQ_0^*(x)dW(x)Q_{i}(x)=\int_\Delta Q_0^*(x)dW(x)Q_0(x)\Phi_{0,i}^{(n)}.
$$
Hence, we obtain the Karlin McGregor Formula for non-symmetric QMCs:
\begin{equation}\label{KmGviaZyg}
\Phi_{0,i}^{(n)}=\left(\int_\Delta dW(x)\right)^{-1}\int_\Delta x^n dW(x)Q_{i}(x),\;\;i\in\mathbb{Z}_{\geq0},\;\;n=0,1,2,\ldots
\end{equation}
This equation gives, for a fixed vertex $i\in\mathbb{Z}_{\geq0},$ the $(0,i)$-th block entry of $\Phi^n$ for any time $n\geq 0.$ The case $n\geq i$
follows from the construction above and, for $n<i,$ $\Phi_{0,i}^{(n)}=0_{d^2}$ since $\Phi$ is block tridiagonal and the right-hand side of
equation \eqref{KmGviaZyg} vanishes by equation \eqref{zygortohogoanlity}. Therefore, we can obtain the probability for the walker to reach site $\ket{0}$,
given that it started on site $\ket{i}$ with initial state $\rho\in M_N(\mathbb{C})$, by
\begin{equation*}\label{probNS}
p_{0i;\rho}(n)=\mathrm{Tr}\left(\Phi_{0,i}^{(n)}\rho\right)=\mathrm{Tr}\left(\left(\int_\Delta dW(x)\right)^{-1}\int_\Delta x^n
dW(x)Q_{i}(x)\rho\right),\;i\in\mathbb{Z}_{\geq0},\;n=0,1,2,\ldots \;.
\end{equation*}
Regarding the case of a finite number of vertices $V=\{0,1,2,\ldots,N\}$, we proceed as expected: the eigenvalues of $\Phi$ are the roots of the determinant of
\begin{equation*}\label{quantumR2NP}
R_{N+1}(x)=Q_N(x)(xI-B_N)-Q_{N-1}(x)C_N,
\end{equation*}
where $\{Q_n(x)\}_{n=0}^N$ are the polynomials associated with $\Phi$. Suppose that $\Phi$ describes a homogeneous QMC, then $\{Q_n(x)\}_{n=0}^N$ are semi-orthogonal with respect to the measure
\begin{equation*}\label{W_kQMCNP}
W_k=\lim_{z\rightarrow \lambda_k}(\lambda_k-z)\left([\Phi]-zI\right)_{00}^{-1},
\end{equation*}
that is,
$$
\sum_{k=1}^{\tau}\lambda_k^iW_kQ_j(\lambda_k)=0,
$$
for $j>i,$ where $\tau$ is the number of eigenvalues of $\Phi$ counting multiplicities. The Karlin-McGregor formula for this kind of QMC is then
\begin{equation*}\label{KMGfiniteQMCNP}
\Phi_{0j}^{(n)}=\sum_{k=1}^{\tau}\lambda_k^n W_kQ_j(\lambda_k).
\end{equation*}

\bex
Let $\Phi$ be the homogeneous OQW with 3 vertices defined by
\begin{equation}\label{ACatt}
\Phi=\begin{bmatrix}
       0 & \lceil C \rceil & 0 \\
       \lceil A \rceil & 0 & \lceil C \rceil \\
       0 &\lceil A \rceil &0
     \end{bmatrix},\quad
A=\frac{1}{\sqrt{3}}\begin{bmatrix}
     1 & 1 \\
     0 & 1
   \end{bmatrix}\quad
C=\frac{1}{\sqrt{3}}\begin{bmatrix}
     1 & 0 \\
     -1 & 1
   \end{bmatrix}.
\end{equation}
The polynomials associated with $\Phi$ are
$$
Q_0(x)=I_4,\;Q_1(x)=x\lceil A \rceil^{-1},\;Q_2(x)=xQ_1(x)\lceil A \rceil^{-1}-\lceil C \rceil \lceil A \rceil^{-1}.
$$
Hence the eigenvalues of $\Phi$ are precisely the roots of
$$
R_3(x)=xQ_2(x)-Q_1(x)\lceil C \rceil ,
$$
which are
\begin{align*}
 \lambda_1&=0,\quad \lambda_2=-\frac{\sqrt{2}}{3},\quad\lambda_3=\frac{\sqrt{2}}{3},\quad\lambda_4=-\frac{\sqrt{3}}{3},\quad\lambda_5=\frac{\sqrt{3}}{3},\\
 \lambda_6&=-\frac{\sqrt{2\sqrt{6} - 3}}{6}+i\frac{\sqrt{2\sqrt{6} + 3}}{6},\quad  \lambda_7=\frac{\sqrt{2\sqrt{6} - 3}}{6}-i\frac{\sqrt{2\sqrt{6} + 3}}{6},\\
  \lambda_8&=-\frac{\sqrt{2\sqrt{6} - 3}}{6}-i\frac{\sqrt{2\sqrt{6} + 3}}{6},\quad  \lambda_9=\frac{\sqrt{2\sqrt{6} - 3}}{6}+i\frac{\sqrt{2\sqrt{6} + 3}}{6}.
\end{align*}
Joining the results of \cite{grunb} and \cite{zyge2}, we obtain
$$
\sum_{k=1}^{9}Q_i^*(\lambda_k)W_kQ_j(\lambda_k)=\begin{cases}
                                                  0_4, & \mbox{if }\quad i>j \\
                                                  F_{ij}\in M_4(\mathbb{C}), \mbox{not necessarily null}& \mbox{ if }\quad i\leq j
                                                \end{cases},
$$
where
\begin{eqnarray}
\nonumber  W_k &=& \lim_{z\rightarrow\lambda_k}(\lambda_k-z)([\Phi]-zI_{12})^{-1}_{00} \\
 \nonumber  &=& \lim_{z\rightarrow\lambda_k}\left((\lambda_k-z)\frac{1}{81z^6-3z^2-2}\times\right.\\
\nonumber&&\left.\begin{bmatrix}
  -\frac{81z^6+9z^4-2z^2-2}{z} & -\frac{27z^4+6z^2-1}{3z} & -\frac{27z^4+6z^2-1}{3z} &-z(9z^2+5) \\
  \frac{27z^4+6z^2-1}{3z} & -\frac{729z^8-162z^6-54z^4-z^2+2}{z(9z^2-2)} & \frac{z(81z^4+27z^2-14)}{9z^2-2} & \frac{21z^2+1}{3z} \\
  \frac{27z^4+6z^2-1}{3z} & \frac{z(81z^4+27z^2-14)}{9z^2-2} & -\frac{729z^8-162z^6-54z^4-z^2+2}{z(9z^2-2)} & \frac{21z^2+1}{3z} \\
  -z(9z^2+5) & -\frac{21z^2+1}{3z}& -\frac{21z^2+1}{3z} & -z(81z^4+7)
\end{bmatrix}\right).
\end{eqnarray}

Those values are
$$
W_1=\frac{1}{6}\begin{bmatrix}
                 6 & 1 & 1 & 0 \\
                 -1 & 3 & 0 & 1 \\
                 -1 & 0 & 3 & 1 \\
                 0 & -1 & -1 & 0
               \end{bmatrix},\;
W_2=W_3=\frac{1}{8}\begin{bmatrix}
                 0 & 0 & 0 & 0 \\
                 0 & 1 & -1 & 0 \\
                 0 & -1 & 1 & 0 \\
                 0 & 0 & 0 & 0
               \end{bmatrix},\;
W_4=W_5=\frac{1}{12}\begin{bmatrix}
                 1 & 1 & 1 & 2 \\
                 -1 & -1 & -1 & -2 \\
                  -1 & -1 & -1 & -2 \\
                 2 & 2 & 2 & 4
               \end{bmatrix},
$$
$$
W_6=W_7=\begin{bmatrix}
        \frac{3-i\sqrt{5}}{-90+6i\sqrt{15}} & -\frac{1}{12}& -\frac{1}{12} & \frac{7-i\sqrt{15}}{-30+18i\sqrt{15}} \\
          -\frac{1}{12} & \frac{5}{30-6i\sqrt{15}} & \frac{5}{30-6i\sqrt{15}} & \frac{-15-7i\sqrt{15}}{-180+12i\sqrt{15}} \\
          -\frac{1}{12} & \frac{5}{30-6i\sqrt{15}} & \frac{5}{30-6i\sqrt{15}} & \frac{-15-7i\sqrt{15}}{-180+12i\sqrt{15}}\\
          \frac{7-i\sqrt{15}}{-30+18i\sqrt{15}} & \frac{15+7i\sqrt{15}}{-180+12i\sqrt{15}} & \frac{15+7i\sqrt{15}}{-180+12i\sqrt{15}} & \frac{11+3i\sqrt{15}}{-30+18i\sqrt{15}}
        \end{bmatrix},
$$
$$
W_8=W_9=\begin{bmatrix}
          \frac{-3-i\sqrt{5}}{-90+6i\sqrt{15}} & -\frac{1}{12}& -\frac{1}{12} & \frac{-7-i\sqrt{15}}{-30+18i\sqrt{15}} \\
          -\frac{1}{12} & -\frac{5}{30-6i\sqrt{15}} & -\frac{5}{30-6i\sqrt{15}} & \frac{15-7i\sqrt{15}}{-180+12i\sqrt{15}} \\
          -\frac{1}{12} & -\frac{5}{30-6i\sqrt{15}} & -\frac{5}{30-6i\sqrt{15}} & \frac{15-7i\sqrt{15}}{-180+12i\sqrt{15}}\\
          \frac{-7-i\sqrt{15}}{-30+18i\sqrt{15}} & \frac{-15+7i\sqrt{15}}{-180+12i\sqrt{15}} & \frac{-15+7i\sqrt{15}}{-180+12i\sqrt{15}} & \frac{-11+3i\sqrt{15}}{30+18i\sqrt{15}}
        \end{bmatrix}.
$$
A simple calculation shows that
$$
dW(x)=\sum_{k=1}^{9}W_k=I_4.
$$
Therefore the Karlin-McGregor formula for this OQW is
$$
\Phi_{0,i}^{(n)}=\left(\int_\Delta dW(x)\right)^{-1}\int_\Delta x^ndW(x)Q_{i}(x)=\sum_{k=1}^{9}\lambda_k^nW_kQ_i(\lambda_k),\quad i=0,1,2,\quad n\geq i.
$$
For instance, we have
$$
\Phi_{0,2}^{(10)}=\sum_{k=1}^{9}\lambda_k^nW_kQ_2(\lambda_k)=\frac{1}{59049}\begin{bmatrix}
                                                                           63 & -45 & -45 & 54 \\
                                                                           -27 & 26 & 10 & -45 \\
                                                                           -27 & 10 & 26 & -45 \\
                                                                           90 & -27 & -27 & 63
                                                                         \end{bmatrix},
$$
which agrees with the corresponding block of $\Phi^{10}.$ The probability of the walker to be on site $\ket{0}$ after 10 steps, given that it started on site $\ket{2}$ with initial density operator
$\rho=\begin{bmatrix}a & b \\b^* & 1-a\end{bmatrix}$ is
$$
p_{02;\rho}(10)=\mathrm{Tr}\left[vec^{-1}\left(\frac{1}{59049}\begin{bmatrix}
                                                                           63 & -45 & -45 & 54 \\
                                                                           -27 & 26 & 10 & -45 \\
                                                                           -27 & 10 & 26 & -45 \\
                                                                           90 & -27 & -27 & 63
                                                                         \end{bmatrix}
\begin{bmatrix}
  a \\
  b \\
  b^* \\
  1-a
\end{bmatrix}\right)\right]=\frac{13+4a-16Re(b)}{6561}.
$$
Analogously,
$$
p_{02;\rho}(2)=\frac{1+4a-4Re(b)}{9},\;\;\;p_{02;\rho}(3)=0,\;\;\;p_{02;\rho}(4)=\frac{1}{27}.
$$
However, the general Karlin-McGregor formula does not apply for this OQW. Indeed, we have
$$
\Phi_{2,2}^{(2)}=\frac{1}{9}\begin{bmatrix}
                     0 & 0 & 0 & 1 \\
                     0 & 0 & -1 & 1 \\
                     0 & -1 & 0 &1 \\
                     1 & -1 & -1 & 1
                   \end{bmatrix},
$$
and
$$
\frac{1}{18}
\begin{bmatrix}
  15 & 37 & 37 & 82 \\
  24& 32 & 30 & 18 \\
  24& 30 & 32 & 18 \\
  25& 29 & 29 & 6
\end{bmatrix}=\left(\sum_{k=1}^{9}Q_2^*(\lambda_k)W_kQ_2(\lambda_k)\right)^{-1}\left(\sum_{k=1}^{9}\lambda_k^2Q_2^*(\lambda_k)W_kQ_2(\lambda_k)\right)\neq\Phi_{2,2}^{(2)}.
$$
The reason why this is happening is that $Q_2$ and $Q_0$ are not orthogonal, since
$$\sum_{k=1}^{9}Q_0^*(\lambda_k)W_kQ_2(\lambda_k)=\frac{1}{4}
\begin{bmatrix}
  -2 & 4 & 4 & 28 \\
  -8 & -21 & -21 & -62 \\
  -8 & -21 & -21 & -62 \\
  4 & 18 & 18 & 68
\end{bmatrix}.$$

\medskip

\begin{figure}[!ht]
\centering
\includegraphics[width=0.8\textwidth]{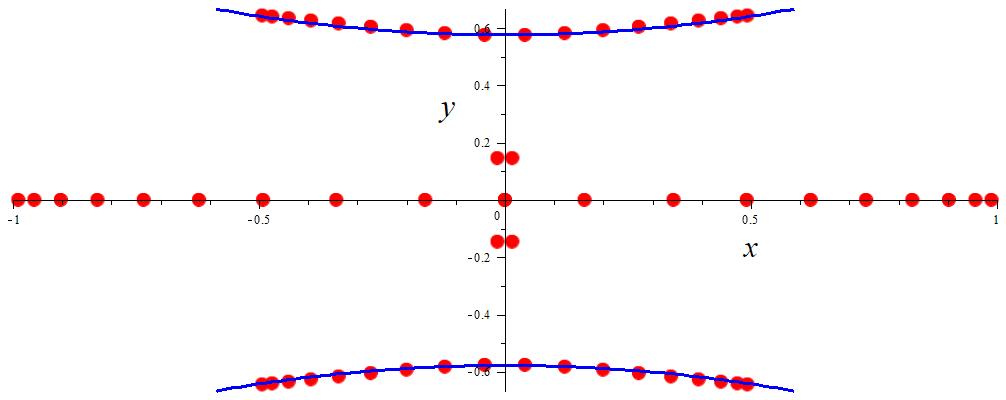}
\caption{Eigenvalues of $\check{\Phi}$ with 20 vertices.}
\label{20vcurve}
\end{figure}

Let us study now the case of a larger number of sites $n$. Consider
$$
\Phi=\begin{bmatrix}
         0 & \lceil C\rceil&  &  &  \\
         \lceil A\rceil& 0 & \lceil C\rceil &  &  \\
         & \ddots & \ddots & \ddots &  \\
          &  &\lceil A\rceil & 0 & \lceil C\rceil \\
          &  &  & \lceil A\rceil & 0
       \end{bmatrix}\in M_{4n}({\mathbb{C}}),
$$
where $A,C$ are defined by \eqref{ACatt}. The compact form of $\Phi$ is given by
$$
\check{\Phi}=\begin{bmatrix}
         0 & C&  &  &  \\
         A & 0 & C &  &  \\
         & \ddots & \ddots & \ddots &  \\
          &  & A & 0 & C \\
          &  &  & A & 0
       \end{bmatrix}\in M_{3n}({\mathbb{C}}),\quad
A=\frac{1}{3}\begin{bmatrix}
    1 & 0 & 0 \\
    -1 & 1 & 0 \\
    1 & -2 & 1
  \end{bmatrix},\quad C=\frac{1}{3}\begin{bmatrix}
    1 & 2 & 1 \\
    0 & 1 & 1 \\
    0 & 0 & 1
  \end{bmatrix}.
$$
If we evaluate the eigenvalues $\lambda_1,\ldots,\lambda_{3n}$ of $\check{\Phi}$ and put them on the complex plane, the outcome is a graph of the form represented in Figure \ref{20vcurve}. Each dot represents an eigenvalue of $\check{\Phi}.$
\eex
\qee

\bex

Let $\Phi$ be a homogeneous QMC with 5 vertices defined by
$$\Phi=
\begin{bmatrix}
  \lceil B_0\rceil &\lceil C_1\rceil +\lceil C_2\rceil  & 0 & 0 & 0 \\
  \lceil A_1\rceil +\lceil A_2\rceil  &\lceil B_0\rceil  & \lceil C_1\rceil +\lceil C_2\rceil  & 0 & 0 \\
   &  \lceil A_1\rceil +\lceil A_2\rceil &  \lceil B_0\rceil & \lceil C_1\rceil +\lceil C_2\rceil & 0 \\
   &  &  \lceil A_1\rceil +\lceil A_2\rceil & \lceil B_0\rceil & \lceil C_1\rceil +\lceil C_2\rceil \\
   &  &  &  \lceil A_1\rceil +\lceil A_2\rceil &  \lceil B_0\rceil
\end{bmatrix},
$$
where
$$
B_0=\frac{\sqrt{5}}{5}\begin{bmatrix}
                      0 & 0 \\
                      0 & 1
                    \end{bmatrix},\;
C_1=\frac{\sqrt{5}}{5}\begin{bmatrix}
                      1 & 0 \\
                      0 & 1
                    \end{bmatrix},\;
C_2=\frac{\sqrt{5}}{5}\begin{bmatrix}
                      0 & 0 \\
                      0 & 1
                    \end{bmatrix},\;
A_1=\frac{\sqrt{5}}{5}\begin{bmatrix}
                      1 & 0 \\
                      -1 & 1
                    \end{bmatrix},\;
A_2=\frac{\sqrt{5}}{5}\begin{bmatrix}
                      1 & 0 \\
                      1 & 1
                    \end{bmatrix}.
$$
In compact form, $\Phi$ becomes
$$\check{\Phi}=
\begin{bmatrix}
  B & C & 0 & 0 & 0 \\
  A & B & C & 0 & 0 \\
  0 & A & B & C & 0 \\
  0 & 0 & A & B & C \\
  0 & 0 & 0 & A & B
\end{bmatrix},\;
B=\frac{1}{5}\begin{bmatrix}
  0 & 0 & 0 \\
  0 & 0 & 0 \\
  0 & 0 & 1
\end{bmatrix},\;
A=\frac{1}{5}\begin{bmatrix}
  2 & 0 & 0 \\
  0 & 2 & 0 \\
  2 & 0 & 2
\end{bmatrix},\;
C=\frac{1}{5}\begin{bmatrix}
  1 & 0 & 0 \\
  0 & 1 & 0 \\
  0 & 0 & 2
\end{bmatrix}.
$$
The eigenvalues of $\check{\Phi}$ are given by
$$
\lambda_1=0,\;\lambda_2=-\frac{1}{5},\;\lambda_3=\frac{1}{5},\;\lambda_4=\frac{3}{5},\;\lambda_5=-\frac{\sqrt{2}}{5},\;\lambda_6=\frac{\sqrt{2}}{5},
$$
$$
\lambda_7=-\frac{\sqrt{6}}{5},\;\lambda_8=\frac{\sqrt{6}}{5},\;\lambda_9=\frac{1}{5}-\frac{2\sqrt{3}}{5},\;\lambda_{10}=\frac{1}{5}+\frac{2\sqrt{3}}{5},
$$
and the weight matrix is given by
$$
W_1=\begin{bmatrix}
  1/3 & 0 & 0 \\
  0 & 1/3 & 0 \\
  2/11 & 0 & 0
\end{bmatrix},\;
W_2=\begin{bmatrix}
  0 & 0 & 0 \\
  0 & 0 & 0 \\
  -1/2 & 0 & 1/4
\end{bmatrix},\;
W_3=\begin{bmatrix}
  0 & 0 & 0 \\
  0 & 0 & 0 \\
  -8/15 & 0 & 1/3
\end{bmatrix},\;
W_4=\begin{bmatrix}
  0 & 0 & 0 \\
  0 & 0 & 0 \\
  1/6 & 0 & 1/4
\end{bmatrix}
$$
$$
W_5=\begin{bmatrix}
  1/4 & 0 & 0 \\
  0 & 1/4 & 0 \\
  \frac{104+\sqrt{2}}{292} & 0 & 0
\end{bmatrix},\;
W_6=\begin{bmatrix}
  1/4 & 0 & 0 \\
  0 & 1/4 & 0 \\
  \frac{104-\sqrt{2}}{292} & 0 & 0
\end{bmatrix},\;
W_7=\begin{bmatrix}
  1/12 & 0 & 0 \\
  0 & 1/12 & 0 \\
  -\frac{17\sqrt{6}}{20}-\frac{67}{30} & 0 & 0
\end{bmatrix},\;
$$
$$
W_8=\begin{bmatrix}
  1/12 & 0 & 0 \\
  0 & 1/12 & 0 \\
  -\frac{17\sqrt{6}}{20}-\frac{67}{30} & 0 & 0
\end{bmatrix},\;
W_9=\begin{bmatrix}
  0 & 0 & 0 \\
  0 & 0 & 0 \\
  \frac{10529}{4818}+\frac{3016\sqrt{3}}{2409} & 0 & 1/12
\end{bmatrix},\;
W_{10}=\begin{bmatrix}
  0 & 0 & 0 \\
  0 & 0 & 0 \\
  \frac{10529}{4818}-\frac{3016\sqrt{3}}{2409} & 0 & 1/12
\end{bmatrix}.
$$
The polynomials $Q_n(x)$ associated with $\check{\Phi}$ (see \eqref{3thermmatrixb}) satisfy \eqref{zygortohogoanlity}, that is,
$$
\sum_{j=1}^{10}\lambda_j^nW(j)Q_k(\lambda_j)=0,
$$
for all integers $n>k\geq 0.$ As an example, formula \eqref{KmGviaZyg} gives, for $\rho=\begin{bmatrix}
                     a & b \\
                     b^* & 1-a
                   \end{bmatrix}$, that
$$
\check{\Phi}_{0,3}^{(7)}=\sum_{k=1}^{10}\lambda_k^7W(k)Q_3(\lambda_k)=\frac{8}{78125}
\begin{bmatrix}
  52 & 0 & 0 \\
  0 & 52 & 0 \\
  907 & 0 & 579
\end{bmatrix}\;\Longrightarrow\; p_{03;\rho}(7)=\frac{4632+608a}{15625}.
$$

\eex\qee

Let us now consider the case of infinite vertices. For that we recall that the Stieltjes transform $B(z;W)$ associated with a homogeneous QMC $\Phi$ with matrix representation
\begin{equation*}\label{phiNPhomo}
\Phi=\begin{bmatrix}
  B & C &  &  & \\
  A & B & C &  &  \\
   & A & B & C &  \\
   &  & \ddots & \ddots &  \ddots
\end{bmatrix},
\end{equation*}
where $A,C\in M_{N^2}(\mathbb{C})$ are non-singular, is given by
\begin{equation}\label{HomoStieltjes}
B(z;W)=(z-B-CB(z;W)A)^{-1}.
\end{equation}
Similarly, the Stieltjes transform $B(z;\widetilde W)$ associated with a QMC $\widetilde\Phi$ with matrix representation
\begin{equation*}\label{phiNPhomonot}
\widetilde\Phi=\begin{bmatrix}
  B_0 & C &  &  & \\
  A_0 & B & C &  &  \\
   & A & B & C &  \\
   &  & \ddots & \ddots &  \ddots
\end{bmatrix},
\end{equation*}
where $A_0,A,C\in M_{N^2}(\mathbb{C})$ are non-singular, is given by
\begin{equation}\label{HomoStieltjesPert}
B(z;\widetilde W)=(z-B_0-CB(z;W)A_0)^{-1}.
\end{equation}

\bex
Take $V=\mathbb{Z}_{\geq0}$ and matrices $R=L=\frac{1}{\sqrt{2}}I_2,$
$$
B_1=\frac{\sqrt{5}}{5}\begin{bmatrix}
                      1 & 0 \\
                      0 & 1
                    \end{bmatrix},\quad
B_2=\frac{\sqrt{5}}{5}\begin{bmatrix}
                      0 & 0 \\
                      0 & 1
                    \end{bmatrix},\quad
R_1=\frac{\sqrt{5}}{5}\begin{bmatrix}
                      1 & 0 \\
                      -1 & 1
                    \end{bmatrix},\quad
R_2=\frac{\sqrt{5}}{5}\begin{bmatrix}
                      1 & 0 \\
                      1 & 1
                    \end{bmatrix}.
$$
We define a QMC on $V$ whose compact form is
$$
\check{\Phi}=\begin{bmatrix}
  B_0 & C &  & & &    \\
  A_0 & 0 & C & & &    \\
 &A & 0 & C & &     \\
  &&A & 0 & C &      \\
   &&  & \ddots & \ddots & \ddots
\end{bmatrix},\;\;B_0=\check{B_1}+\check{B_2},\;A_0=\check{R_1}+\check{R_2},\;C=\check{L},\;A=\check{R}.
$$
Denote by $\check{\Phi}_0$ the matrix
$$\check{\Phi}_0=\begin{bmatrix}
   0 & C &  & &   \\
 A & 0 & C &&      \\
 &A & 0 & C &      \\
    & & \ddots & \ddots & \ddots
\end{bmatrix},$$
and $W, W_0$ the weight matrices associated with $\check{\Phi}$  and $\check{\Phi}_0,$ respectively. Using \eqref{HomoStieltjes} and \eqref{HomoStieltjesPert} we obtain
$$
B(z;W_0)(z)=(2z+2\sqrt{z^2-1})I_3.
$$
and
$$
B(z;W)=\frac{5}{5z^2-6z+5}
\begin{bmatrix}
  2\sqrt{z^2-1}+3z-1 & 0 & 0 \\
  0 & 2\sqrt{z^2-1}+3z-1 & 0 \\
  \frac{2\left((25z^2-20z-1)\sqrt{z^2-1}+25z^3-20z^2-13z+8\right)}{5z^2-18z+13} & 0 & 2\sqrt{z^2-1}+3z-3
\end{bmatrix}.
$$
With the Stieltjes transform, we may obtain the associated weight matrix for $\check{\Phi}$ by applying the Perron-Stieltjes inversion formula. A simple calculation shows that the weight matrix $W$ is given by
$$
W(x)=\frac{5}{\pi(5x^2-6x+5)}
\begin{bmatrix}
  2\sqrt{1-x^2} & 0 & 0 \\
  0 & 2\sqrt{1-x^2} & 0 \\
 \displaystyle \frac{2(25x^2-20x-1)\sqrt{1-x^2}}{5x^2-18x+13} & 0 & 2\sqrt{1-x^2}
\end{bmatrix},\quad x\in[-1,1].
$$
We now have
$$
\int_{-1}^{1}Q_i^*(x)dW(x)Q_j(x)=0,\quad i>j,
$$
thus formula \eqref{KmGviaZyg} holds.

\medskip

Let us now analyze recurrence of the first vertex of both QMCs $\check{\Phi}$ and $\check{\Phi}_0.$ By \eqref{corqmcstieltjes}, we are able to conclude whether the walk is recurrent just by considering the Stieltjes transform associated with the QMC, that is, we do not need to obtain the explicit weight matrix associated with the referred QMC. Above, we determined the weight matrix for completeness, and in order to write the transitions probabilities of the walk described by $\Phi$ using the Karlin-McGregor formula.

\medskip

Applying limits to the Stieltjes transform $B(z;W_0)$ and $B(z;W)$ associated with $\check{\Phi}_0$ and $\check{\Phi},$ respectively, we obtain
$$
\displaystyle\lim_{z\rightarrow 1}\mathrm{Tr}(B(z,W_0)\rho)=\lim_{z\rightarrow 1}2z+2\sqrt{z^2-1}=2,$$
and using l'Hospital's rule we get
$$
\lim_{z\rightarrow 1}\mathrm{Tr}(B(z,W)\rho)=\infty,
$$
for any density operator $\rho\in M_2(\mathbb{C}).$ Therefore, by \eqref{corqmcstieltjes}, the first vertex $\ket{0}$ is transient for $\check{\Phi}_0$ and recurrent for $\check{\Phi}.$
\eex
\qee

\bex\label{exNPZ+}
Take $V=\mathbb{Z}_{\geq0}$ and matrices
\begin{equation}\label{R1R2L1}
R_1=\frac{1}{\sqrt{7}}\begin{bmatrix}
                      1 & 0 \\
                      -1 & \sqrt{3}
                    \end{bmatrix},\quad
R_2=\frac{1}{\sqrt{7}}\begin{bmatrix}
                      1 & 0 \\
                      1 & \sqrt{3}
                    \end{bmatrix},\quad
L_1=\frac{1}{\sqrt{7}}\begin{bmatrix}
                      \sqrt{3} & 0 \\
                      0 & 1
                    \end{bmatrix}.
\end{equation}
We define a QMC on $V$ whose compact form is
$$
\check{\Phi}=\begin{bmatrix}
  0 & C &  &  &    \\
  A & 0 & C &  &    \\
 &A & 0 & C &      \\
   &  & \ddots & \ddots & \ddots
\end{bmatrix},\;\;A=\check{R_1}+\check{R_2},\;C=\check{L_1}.
$$
The Stieltjes transform associated with $\check{\Phi}$ satisfies
$$
B(z;W)(zI_3-CB(z;W)A)=I_3,
$$
for which a solution is
\begin{equation}\label{fafsag}
B(z;W)=\frac{7}{12}
\begin{bmatrix}
  7z-i\sqrt{-49z^2+24} & 0 & 0 \\
  0 & 7z-i\sqrt{-49z^2+24} & 0 \\
  \displaystyle\frac{-343z^3+140z+(49z^2-8)\sqrt{49z^2-24}}{49z^2-32} & 0 & 7z-i\sqrt{-49z^2+24}
\end{bmatrix}.
\end{equation}
The weight matrix associated with $\check{\Phi}$ is then
$$
W(x)=\frac{7}{12}
\begin{bmatrix}
  \sqrt{24-49x^2} & 0 & 0 \\
  0 & \sqrt{24-49x^2} & 0 \\
  -\displaystyle\frac{(49x^2+8)\sqrt{24-49x^2}}{49x^2-32} & 0 & \sqrt{24-49x^2}
\end{bmatrix},\quad x\in\left[-\frac{2\sqrt{6}}{7},\frac{2\sqrt{6}}{7}\right].
$$
The polynomials associated with $\check{\Phi},$ $Q_k(x),$ satisfy
$$
\int_{-\frac{2\sqrt{6}}{7}}^{\frac{2\sqrt{6}}{7}}x^idW(x)Q_j(x)=0,\quad i>j,
$$
thus formula \eqref{KmGviaZyg} holds. Finally, we conclude that vertex $\ket{0}$ is transient, since
\begin{align*}
\sum_{n=0}^{\infty}p_{00;\rho}(n) &= \lim_{z\rightarrow 1}\mathrm{Tr}\left(B(z,W)\rho\right) \\
&= \frac{49z-7\sqrt{49z^2-24}}{12}+\frac{7a}{12}\frac{-343z^3+140z+(49z^2-8)\sqrt{49z^2-24}}{49z^2-32}= \frac{119+7a}{102}<\infty.
\end{align*}
\eex
\qee

\bex Let us consider the QMC on $V=\mathbb{Z}_{\geq0}$ whose compact form is
$$
\check{\Phi}=\begin{bmatrix}
  C & C &  &  &    \\
  A & 0 & C &  &    \\
 &A & 0 & C &      \\
   &  & \ddots & \ddots & \ddots
\end{bmatrix},\;\;A=\check{R_1}+\check{R_2},\;C=\check{L_1},
$$
where
$$
R_1=\frac{1}{\sqrt{7}}\begin{bmatrix}
                      1 & 0 \\
                      -1 & \sqrt{3}
                    \end{bmatrix},\quad
R_2=\frac{1}{\sqrt{7}}\begin{bmatrix}
                      1 & 0 \\
                      1 & \sqrt{3}
                    \end{bmatrix},\quad
L_1=\frac{1}{\sqrt{7}}\begin{bmatrix}
                      \sqrt{3} & 0 \\
                      0 & 1
                    \end{bmatrix}.
$$
This QMC is similar to the one on Example \ref{exNPZ+} with the difference that the first block is replaced by $C.$ Now $\check{\Phi}$ is trace preserving and the associated Stieltjes transform to $\check{\Phi},$  $B(z;W),$ satisfies
$$
B(z;W)(zI_3-C-CB(z;\tilde{W})A)=I_3,
$$
where $B(z;\tilde{W})$ is the associated Stieltjes transform to the QMC on Example \ref{exNPZ+}.
Thus, we obtain
$$
B(z;W)=
\begin{bmatrix}
  \frac{7}{6}\frac{7z-6+\sqrt{49z^2-24}}{5-7z} & 0 & 0 \\
  0 &  \frac{7}{2}\frac{-7z+2\sqrt{3}-\sqrt{49z^2-24}}{7\sqrt{3}z-9} & 0 \\
  \frac{343z^3-196z^2-126z+64+(49z^2-28z-4)\sqrt{49z^2-24}}{160-384z-21z^2+588z^3-343z^4} & 0 &  \frac{1}{2}\frac{7z-2+\sqrt{49z^2-24}}{1-z}
\end{bmatrix}.
$$
Therefore,
\begin{align*}
 \sum_{n=0}^{\infty}p_{00;\rho}(n) &= \lim_{z\rightarrow 1}\mathrm{Tr}\left(B(z;W)\rho\right) \\
   &= \frac{7}{3}\frac{(343z^3+(49z^2-20)\sqrt{49z^2-24}-182z)a}{343z^3-245z^2-224z+160}+\frac{1}{2}\frac{7z-2+\sqrt{49z^2-24}}{1-z}=\infty,
\end{align*}
for any density operator $\rho=\begin{bmatrix}
                                                                                       a & b \\
                                                                                       b^* & 1-a
                                                                                     \end{bmatrix}.$
Hence, this QMC is recurrent.
\eex
\qee

{\bf Applying the folding trick to a nonpositive measure.} It is worth noting that the folding trick can also be applied to homogeneous QMCs whose matrix representations are not symmetrizable. Then, we can examine the associated recurrence problem. In fact, let us recall equation \eqref{PP1}:
$$\Phi_{00}(z)=\Phi_{00}^+(z)(I-z^2A_{-1}\Phi_{-1,-1}^-(z)C_0\Phi_{00}^+(z))^{-1}.
$$
In order to analyze recurrence of site $\ket{0}$ of a given QMC on $\mathbb{Z}$, we have to calculate $\sum_{n=0}^{\infty}p_{00;\rho}(n)=\sum_{n=0}^{\infty}\mathrm{Tr}(\Phi_{00}^{(n)}\rho)$ for each density operator $\rho.$ This can be done by using equation \eqref{PP1} in the following way:
\begin{equation}\label{notStiltjes}
\sum_{n=0}^{\infty}\Phi_{00}^{(n)}=\lim_{z\uparrow 1}\Phi_{00}(z)=\lim_{z\uparrow 1}\Pi_{0}^+B(z;W_+)(I-A_{-1}\Pi_{-1}^-B(z;W_-)C_0\Pi_{0}^+B(z;W_+))^{-1},
\end{equation}
where the Stieltjes transform appearing on the right-hand side are obtained by applying \eqref{fubstieltjes}. Therefore, we have the following result, which gives a recurrence criterion for a tridiagonal homogeneous QMC with non-singular coins above and below the main diagonal.
\begin{pro}\label{teoHOMO}
Fix $N \in\{1,2,3,\ldots\},$ $A,B,C$ operators on $\mathbb{C}^{N^2}$ with $A,C$ non-singular such that
$$
\Phi=\begin{bmatrix}
             \ddots & \ddots & \ddots && & &  \\
            & A & B & C &  && \\
             & & A & B & C &&  \\
            & & & A & B & C  & \\
            & & &  & \ddots & \ddots & \ddots
           \end{bmatrix},
$$
is a QMC on $\mathbb{Z}$. Given a density operator $\rho\in M_{N}(\mathbb{C}),$ a vertex $i\in \mathbb{Z}$ is $\rho$-recurrent if and only if
\begin{equation}\label{teoHOMOrec}
\lim_{z\uparrow 1}\mathrm{Tr}\left[B(z;W_+)(I-A\Pi_{0}^+B(z;W_+)C\Pi_{0}^+B(z;W_+))^{-1}\rho\right]=\infty,
\end{equation}
where $B(z;W_+)$ is the solution of \eqref{HomoStieltjes}. Therefore a QMC $\Phi$ is recurrent if and only if \eqref{teoHOMOrec} is satisfied for any density operator $\rho\in M_{N}(\mathbb{C}).$

\begin{proof}
Vertex $\ket{0}$ is $\rho$-recurrent if and only if
$$
\sum_{n=0}^{\infty}\mathrm{Tr}\left(\Phi_{00}^{(n)}\rho\right)=\infty.
$$
Since the QMC is homogeneous we have $\Phi^+=\Phi^-,$ hence \eqref{notStiltjes} gives the equivalence between recurrence and equation \eqref{teoHOMOrec}.
\end{proof}
\end{pro}

\bex
We will extend the QMC on $\mathbb{Z}_{\geq0}$ given by Example \ref{exNPZ+} to $\mathbb{Z}.$ Let $\Phi$ be a homogeneous QMC with compact matrix representation given by
$$
\check{\Phi}=\begin{bmatrix}
\ddots& \ddots & \ddots & & &  &    \\
&A& 0 & C &  &  &    \\
 &&A& 0 & C &  &      \\
  &&&A & 0 & C &      \\
   &&&  & \ddots & \ddots & \ddots
\end{bmatrix},\;\;A=\check{R_1}+\check{R_2},\;C=\check{L_1},
$$
where $R_1,R_2$ and $L_1$ are given by \eqref{R1R2L1}. The Stieltjes transform associated with $\Phi^+$ is the same as the one given by \eqref{fafsag}. Therefore, according to Proposition \ref{teoHOMO}, we have
\begin{eqnarray}
\nonumber  \sum_{n=0}^{\infty}p_{00;\rho}(n) &=& \lim_{z\rightarrow 1}\mathrm{Tr}(B(z;W_+)(I-A\Pi_{0}^+B(z;W_+)C\Pi_{0}^+B(z;W_+))^{-1}\rho) \\
\nonumber   &=& \lim_{z\rightarrow 1}\mathrm{Tr}\left(7
\begin{bmatrix}
  \frac{1}{\sqrt{49z^2-24}} & 0 & 0 \\
  0 & \frac{1}{\sqrt{49z^2-24}} & 0 \\
  \frac{-343z^3+84z+(49z^2+8)\sqrt{49z^2-24}}{49z^2-32} & 0 & \frac{1}{\sqrt{49z^2-24}}
\end{bmatrix}
\begin{bmatrix}
  a \\
  2b \\
  1-a
\end{bmatrix}\right)\\
\nonumber   &=&\lim_{z\rightarrow 1}\mathrm{Tr}\left(
\begin{bmatrix}
                  \frac{7a}{\sqrt{49z^2-24}} \\
                  \frac{14b}{\sqrt{49z^2-24}} \\
                 7a\frac{-343z^3+84z+(49z^2+8)\sqrt{49z^2-24}}{49z^2-32}+\frac{7(1-a)}{\sqrt{49z^2-24}}
\end{bmatrix}\right)\\
\nonumber   &=& \lim_{z\rightarrow 1}7a\frac{-343z^3+84z+(49z^2+8)\sqrt{49z^2-24}}{49z^2-32}+\frac{7}{\sqrt{49z^2-24}}\\
\nonumber   &=&\frac{182a+595}{425},
\end{eqnarray}
for any density operator $\rho=\begin{bmatrix}a & b \\b^* & 1-a\end{bmatrix}.$
Therefore, we conclude that this QMC is transient.

\eex\qee

\medskip

{\bf Acknowledgements.} The work of MDI was partially supported by PAPIIT-DGAPA-UNAM grant IN104219 (M\'exico) and CONACYT grant A1-S-16202 (M\'exico). CFL is grateful for the financial support and hospitality of the Instituto de Matem\'aticas regarding his visit in January 2020, where part of this work was carried out. NL acknowledges financial support by CAPES (Coordena\c c\~ao de Aperfei\c coamento de Pessoal de N\'ivel Superior) during the period 2018-2021.

\end{document}